\documentclass[fleqn,usenatbib,useAMS]{mnras}

\usepackage{newtxtext,newtxmath}
\usepackage[T1]{fontenc}
\usepackage{ae,aecompl}
\usepackage{graphicx}	
\usepackage{amsmath}	
\usepackage{xspace}

\usepackage{xcolor}

\newlength\q
\usepackage{array}
\newcolumntype{C}[1]{>{\centering\let\newline\\\arraybackslash\hspace{0pt}}m{#1}}

\newcommand{\be}{\begin{equation}}
\newcommand{\ee}{\end{equation}}
\newcommand{\bea}{\begin{eqnarray}}
\newcommand{\eea}{\end{eqnarray}}
\newcommand{\no}{\nonumber}

\newcommand{\TNG}{TNG300\xspace}
\newcommand{\Mpeak}{\ensuremath{M_{\rm peak}}\xspace}
\newcommand{\Mpeakz}{\ensuremath{M_{\rm peak}^{z=0}}\xspace}
\newcommand{\tform}{\ensuremath{t_{1/2}}\xspace}
\newcommand{\slope}{\ensuremath{\beta}\xspace}
\newcommand{\Mpeakzall}{\ensuremath{M_{\rm peak}^{z}}\xspace}

\newcommand{\Msun}{\ensuremath{M_\odot}\xspace}

\definecolor{mypink2}{RGB}{219, 48, 122}

\title[The influence of MAH on galaxy properties and assembly bias]
{On the influence of halo mass accretion history on galaxy properties and assembly bias}

\author[Montero-Dorta et al.]{
\parbox[t]{\textwidth}{
Antonio D. Montero-Dorta$^{1}$\thanks{E-mail: amonterodorta@gmail.com}, Jon\'as Chaves-Montero$^{2,3}$, M. Celeste Artale$^{4}$, Ginevra Favole$^{5}$} 
\vspace*{6pt} \\ 
$^1$Departamento de F\'isica, Universidad T\'ecnica Federico Santa Mar\'ia, Casilla 110-V, Avda. Espa\~na 1680, Valpara\'iso, Chile.\\
$^2$ Donostia International Physics Centre, Paseo Manuel de Lardizabal 4, 20018 Donostia-San Sebastian, Spain. \\
$^3$ HEP Division, Argonne National Laboratory, 9700 South Cass Avenue, Lemont, IL 60439, USA. \\
$^4$ Institut f\"ur Astro- und Teilchenphysik, Universit\"at Innsbruck, Technikerstrasse 25/8, 6020 Innsbruck, Austria. \\
$^5$Institute of Physics, Laboratory of Astrophysics, Ecole Polytechnique F\'ed\'erale de Lausanne (EPFL), Observatoire de Sauverny, 1290 Versoix, Switzerland.\\
\vspace{-0.4cm} 
}

\date{Accepted ---. Received ---;in original form --- \vspace{-0.3cm}}

\pubyear{2021}

\def\simlt{\lower.5ex\hbox{$\; \buildrel < \over \sim \;$}}
\def\simgt{\lower.5ex\hbox{$\; \buildrel > \over \sim \;$}}

\definecolor{red}{rgb}{1,0,0}

\begin{document}
\label{firstpage}
\pagerange{\pageref{firstpage}--\pageref{lastpage}}
\maketitle


\begin{abstract}

Halo assembly bias is the secondary dependence of the clustering of dark-matter haloes on their assembly histories at fixed halo mass. This established dependence is expected to manifest itself on the clustering of the galaxy population, a potential effect commonly known as galaxy assembly bias. Using the IllustrisTNG300 magnetohydrodynamical simulation, we analyse the dependence of the properties and clustering of galaxies on the shape of the specific mass accretion history of their hosting haloes (sMAH). We first show that several halo and galaxy properties strongly correlate with the slope of the sMAH (\slope) at fixed halo mass. Namely, haloes with increasingly steeper \slope increment their halo masses faster at early times, and their hosted galaxies present larger stellar-to-halo mass ratios, lose their gas faster, reach the peak of their star formation histories at higher redshift, and become quenched earlier. We also demonstrate that \slope is more directly connected to these key galaxy formation properties than other broadly employed halo proxies, such as formation time. Finally, we measure the secondary dependence of galaxy clustering on \slope at fixed halo mass as a function of redshift. By tracing back the evolution of individual haloes, we show that the amplitude of the galaxy assembly bias signal for the progenitors of $z=0$ galaxies increases with redshift, reaching a factor of 2 at $z = 1$ for haloes of $M_\mathrm{halo}=10^{11.5}-10^{12}$ $h^{-1}\mathrm{M}_\odot$. The measurement of the evolution of assembly bias {\it{along the merger tree}} provides a new theoretical perspective to the study of secondary bias. Our findings, which show a tight relationship between halo accretion and both the clustering and some of the main observational properties of the galaxy population, have also important implications for the generation of mock catalogues for upcoming cosmological surveys.

\end{abstract}

\begin{keywords}

methods: numerical - galaxies: formation - galaxies: haloes - dark matter - large-scale structure of Universe - cosmology: theory

\end{keywords}

\section{Introduction}
\label{sec:intro}

The mass accretion history (MAH) of dark-matter (DM) haloes is the fundamental property regulating the formation and evolution of galaxies in multiple galaxy formation frameworks, from classical semi-analytic models \citep[e.g.,][]{Kauffmann1993, baugh1996_sam, avila1998_sam} to state-of-the-art empirical models \citep[e.g.,][]{becker2015_eam, moster2018_emerge, behroozi2019_um}. This assumption is based on the well-established result that more massive haloes tend to host more massive galaxies, with only relatively small scatter in their stellar mass--halo mass relation (e.g., \citealt{White1978, More2009,Yang2009,Reddick2013,Watson2013,Tinker2013,Gu2016}).

However, due to the complexity of halo evolution, simple models resort to predicting galaxy properties using only some characteristic features of their assembly histories. For example, the mass or potential-well depth of haloes is known to present substantial correlation with galaxy stellar mass \citep{ChavesMontero2016, Feldmann2019}, but cannot predict specific star formation rates (SFRs) or colours \citep{feldmann2017_ColoursStarformation, Feldmann2019}. Conversely, these galaxy properties are directly linked to halo formation time \citep{Behroozi2015, Feldmann2019}. It appears clear, therefore, that none of these basic features can fully capture the dependence of galaxy properties on halo assembly \citep{ChavesMontero2016, Campbell2018_sham, MonteroDorta2020B,Hadzhiyska2021_sham,Favole2021}, and some of them are in fact affected by transient effects \citep{ChavesMontero2016, Wang2020_concentration} and/or subhalo identification and tracking issues \citep{knebe2011_comparison, Knebe2013_comparison, Behroozi2013_grav}. 

In this context, hydrodynamical simulations follow a different approach. They employ known physics to simulate, at a sub-grid level, a variety of processes that are related to galaxy formation, including star formation, radiative metal cooling, and supernova, stellar, and black hole feedback (see reviews in \citealt{Somerville2015,Naab2017}). Importantly, they provide an additional means of testing the validity of empirical models of galaxy formation. In this paper, we use the IllustrisTNG\footnote{\url{http://www.tng-project.org}} magnetohydrodynamical simulation to investigate the dependence of multiple galaxy properties, including SFR and the stellar-to-halo mass ratio, on the MAH of their host haloes.

This MAH analysis is also relevant for the study of the large scale structure. It is well established that the {\it{large-scale}} linear bias of DM haloes\footnote{Throughout this work, by ``large-scale" linear bias we mean the ratio between the clustering of DM haloes and the underlying matter density field on scales 5-15 $h^{-1}$Mpc.} depends strongly on the internal properties of haloes. Among these properties, halo mass is responsible for the primary dependence: more massive haloes are more tightly clustered than less massive haloes, as expected from the $\Lambda$-cold dark matter ($\Lambda$-CDM) structure formation formalism (e.g., \citealt{Press1974,ShethTormen2002}). More recently, a number of additional {\it{secondary dependencies}} at fixed halo mass have been unveiled (see, e.g., \citealt{Sheth2004,gao2005,wechsler2006,Gao2007,Angulo2008,2008Li,Lazeyras2017,2018Salcedo,han2018,Mao2018, SatoPolito2019, Johnson2019, Ramakrishnan2019,MonteroDorta2020B, Tucci2020}). Among these dependencies, the one that has drawn more attention is precisely the dependence on the assembly history of haloes, an effect dubbed {\it{halo assembly bias}}. Typically lower mass haloes that assemble a significant portion of their mass early on are more tightly clustered than haloes that form at later times, with the signal progressively vanishing towards the high-mass end (e.g., \citealt{gao2005, SatoPolito2019}). 

The aforementioned general trend has been shown in simulations to be already in place at early times, even though the amplitude of the halo assembly bias signal decreases with redshift for a given halo mass. This redshift evolution is simply a natural consequence of the principal dependence of the signal on the peak height of fluctuations \citep{Gao2007}. Halo assembly bias is usually measured in terms of the half-mass formation redshift $z_{1/2}$ in $N$-body numerical simulations, i.e. the redshift at which half of the halo peak or present-day mass is formed. It is however known that the amplitude of the signal depends strongly on the adopted definition of halo age \citep{Chue2018}. 

Halo assembly bias is likely a manifestation of more fundamental physical mechanisms connected to large-scale environments. Although the origins of the low mass trend are not established, several theories have attempted to relate the effect with the truncation of mass accretion in a subpopulation of haloes. This ``stalled evolution" could be caused by tidal interactions with nearby haloes (e.g., \citealt{Dalal2008,2018Salcedo}) or by the large-scale tidal fields  (e.g., \citealt{Borzyszkowski2017,Musso2018, Ramakrishnan2019}).  

In the second part of this paper, we investigate the effect called {\it{galaxy assembly bias}}, which is defined throughout this work as a direct manifestation of halo assembly bias on the clustering of the galaxy population - namely, the dependence of galaxy clustering on the MAH at fixed halo mass. Furthermore, we present a novel theoretical perspective where galaxy assembly bias is measured for the progenitors of present-day galaxies. Note that galaxy assembly bias can also be defined globally, as the combined effect of all possible secondary dependencies of galaxy clustering at fixed halo mass (i.e., not just the particular dependence on halo assembly history). We avoid this definition for simplicity. It can also be viewed from the related perspective of halo occupations, i.e., as the dependence of the galaxy content of haloes ({\it{occupancy variations}}) on halo properties beyond halo mass (see, e.g., \citealt{Artale2018,Zehavi2018,Bose2019,Salcedo2020}). Other definitions include the recently proposed {\it{anisotropic assembly bias}}, which employs higher order moments of the power spectrum \citep{Obuljen2019,Obuljen2020}.

Despite several detection claims, galaxy assembly bias is still generally regarded as an unconfirmed hypothesis (see, e.g.,  \citealt{Miyatake2016,Lin2016,Sunayama2016,Zu2016,MonteroDorta2017B,Niemiec2018,Zentner2019,Walsh2019,Sunayama2019,Obuljen2019,MonteroDorta2020C,Obuljen2020,Salcedo2020} for more information). In the context of hydrodynamical simulations, \cite{MonteroDorta2020B} showed that the halo assembly bias signal is present in IllustrisTNG and propagates to the central galaxy population when this is split by properties such as colour or SFR. In this paper, we extend this analysis by measuring galaxy assembly bias in IllustrisTNG with respect to the parametrized shape of the MAH as a function of redshift. 

The analysis of the MAH and the evolution of galaxy assembly bias have important implications for the modelling of the halo-galaxy connection, the construction of galaxy mocks for next-generation cosmological surveys, and the potential detection of the signal at high redshift (see \citealt{Wechsler2018} for a review). These challenges motivate this paper, which is organised as follows. Section~\ref{sec:sims} provides brief descriptions of the simulation box and the halo and galaxy properties analysed in this work. Section~\ref{sec:model} presents our parametrization of the MAH and investigates the relationship between the MAH and several galaxy-formation related properties. The galaxy assembly bias measurements and the  redshift evolution of the signals are shown and discussed in 
Section~\ref{sec:results}. Finally, Section~\ref{sec:discussion} is devoted to discussing the implications of our results and providing a brief summary of the paper. The IllustrisTNG300 simulation adopts the standard $\Lambda$CDM cosmology \citep{Planck2016}, with parameters $\Omega_{\rm m} = 0.3089$,  $\Omega_{\rm b} = 0.0486$, $\Omega_\Lambda = 0.6911$, $H_0 = 100\,h\, {\rm km\, s^{-1}Mpc^{-1}}$ with $h=0.6774$, $\sigma_8 = 0.8159$, and $n_s = 0.9667$.

\begin{figure*}
    \centering
    \includegraphics[width=0.33\textwidth]{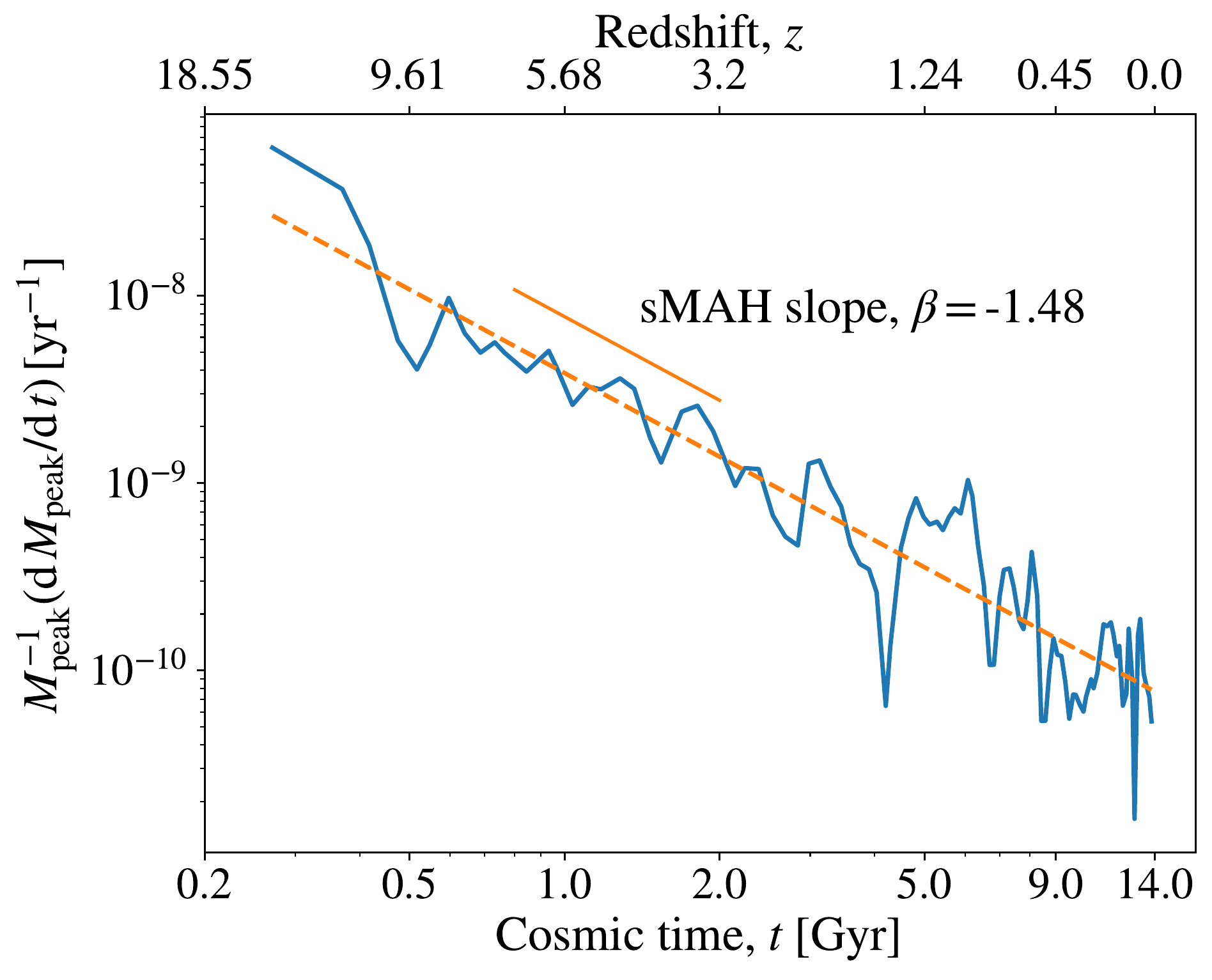}
    \includegraphics[width=0.33\textwidth]{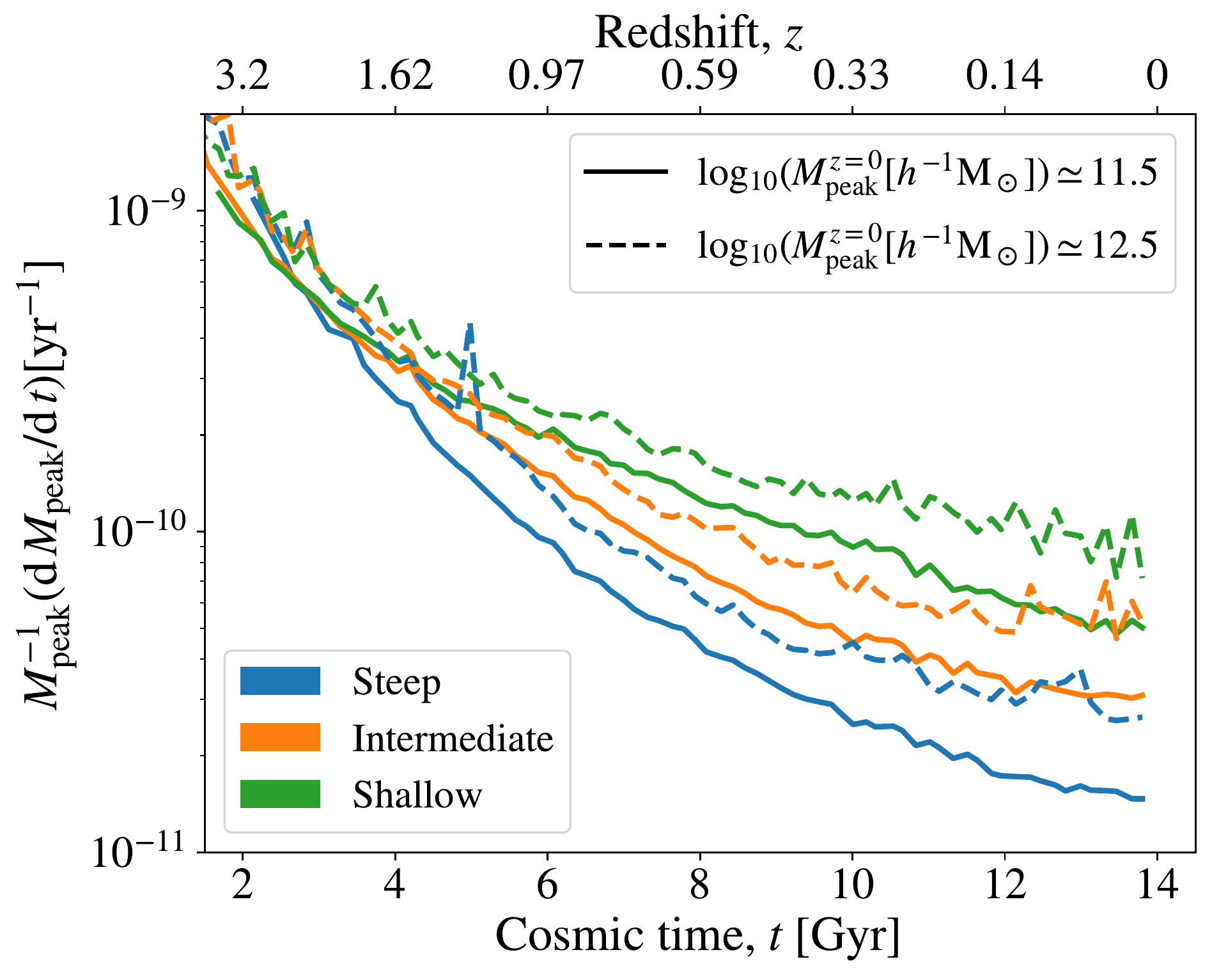}
    \includegraphics[width=0.33\textwidth]{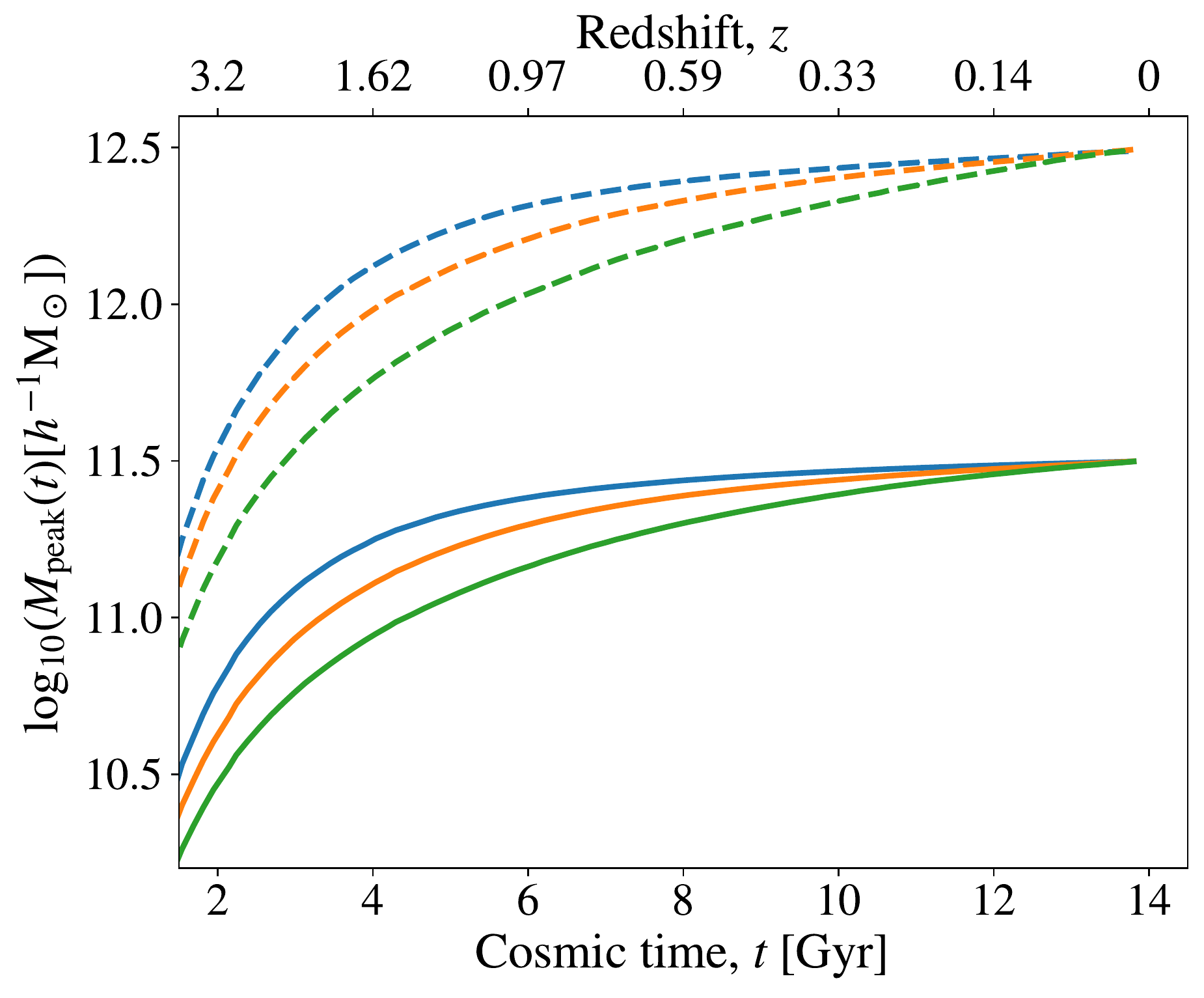}
    
    \caption{{\it Left panel.} Specific mass accretion history (sMAH) for a randomly selected \TNG galaxy. The blue solid line shows the sMAH as predicted by \TNG, while the orange dashed line represents the best-fitting power-law model (see text). The TNG300 sMAHs are generally well described by power laws when displayed as a function of the logarithmic of cosmic time. {\it Middle and right panels}. Average sMAHs and halo mass histories for \TNG galaxies. Solid and dashed lines show results for galaxies hosted by haloes of  $\log_{10}(\Mpeakz [h^{-1}{\rm M_{\odot}}])=11.5$ and 12.5, respectively, while the colour code indicates the value of \slope (the slope of the sMAH). Blue, orange, and green colours represent results for galaxies within the first, second, and third tercile of the \slope distribution at fixed mass. Haloes with increasingly steeper sMAH slope present more (less) accretion at early (late) times.}
    \label{fig:model_sMAH_cartoon}
\end{figure*}


\section{Simulation data}
\label{sec:sims}

Our analysis is based on data from the IllustrisTNG magnetohydrodynamical cosmological simulation \citep{Pillepich2018b,Pillepich2018,Nelson2018_ColorBim,Nelson2019,Marinacci2018,Naiman2018,Springel2018}. The IllustrisTNG simulation suite was produced using the {\sc arepo} moving-mesh code \citep{Springel2010} and is considered an improved version of the previous Illustris simulation \citep{Vogelsberger2014a, Vogelsberger2014b, Genel2014}. The updated IllustrisTNG sub-grid models account for star formation, radiative metal cooling, chemical enrichment from SNII, SNIa, and AGB stars, stellar feedback, and super-massive black hole feedback. These models were calibrated to reproduce a set of observational constraints that include the observed $z=0$ galaxy stellar mass function, the cosmic SFR density, the halo gas fraction, the galaxy stellar size distributions, and the black hole -- galaxy mass relation.

Since we are interested in measuring large-scale halo/galaxy clustering, we chose to analyse the largest box available in the database, IllustrisTNG300-1 (hereafter TNG300). TNG300 spans a side length of $205\,\,h^{-1}$Mpc and includes periodic boundary conditions. The TNG300 run followed the dynamical evolution of 2500$^3$ DM particles of mass $4.0 \times 10^7$ $h^{-1} {\rm M_{\odot}}$ and (initially) 2500$^3$ gas cells of mass $7.6 \times 10^6$ $h^{-1} {\rm M_{\odot}}$. This box is a useful tool for galaxy formation and clustering science that has proven capable of reproducing a number of observational measurements
(see, e.g., \citealt{Springel2018,Pillepich2018,Bose2019,Beltz-Mohrmann2020,Contreras2020,Gu2020,Hadzhiyska2020,Hadzhiyska2021,Shi2020,MonteroDorta2020B,MonteroDorta2020C,Favole2021}).

DM haloes in IllustrisTNG are identified using a friends-of-friends (FOF) algorithm with a linking length of 0.2 times the mean inter-particle separation \citep{Davis1985}. The gravitationally bound substructures that we call subhaloes are in turn identified using the {\sc subfind} algorithm \citep{Springel2001,Dolag2009}. Subhaloes containing a non-zero stellar mass component are labelled galaxies.

Since the cornerstone of this work is the MAH, we make use of the subhalo merger tree catalogues available in the database. The IllustrisTNG database provides merger trees computed with both {\sc SubLink} \citep{Rodriguez-Gomez2015} and {\sc LHaloTree} \citep{Springel2005}. These two algorithms, according to \citealt{Nelson2018}, converge to similar results in a ``population-average sense". Here we opt to use the merger trees from SubLink \citep{Rodriguez-Gomez2015} in order to select and follow back the main branch of subhaloes at $z=0$. 

In our analysis, we use subhalo and halo properties directly extracted from the database. These properties are instrumental in characterising subhaloes and their MAHs, along with the impact that the latter have on the galaxy population. For haloes, we use the virial mass, $M_{\rm vir}$ [$h^{-1} {\rm M_{\odot}}$], defined simply as the total mass enclosed within a sphere of radius $R_{\rm vir}$, where the density equals 200 times the critical density. From $M_{\rm vir}$ and the merger trees provided in the database, we compute both the redshift-dependent peak mass, \Mpeakzall, and the half-peak-mass formation time \tform (i.e., the time at which half the $z=0$ peak mass \Mpeakz has formed).

For the simulated galaxies, we use the stellar mass, $M_\ast{}$ [$h^{-1} {\rm M_{\odot}}$], and the gas mass, $M_{\rm gas}$ [$h^{-1} {\rm M_{\odot}}$], computed, respectively, as the sum of the mass of all stellar particles and gas cells bound to each subhalo. These quantities are used to measure the SFR and the star formation history (SFH) of each galaxy.

\begin{figure*}

    \centering
    \begin{tabular}{c}
        \textbf{\Large Slope of the sMAH (\slope)}\\
    \end{tabular}
    \hspace*{0.75cm}
    
    \includegraphics[width=0.33\textwidth]{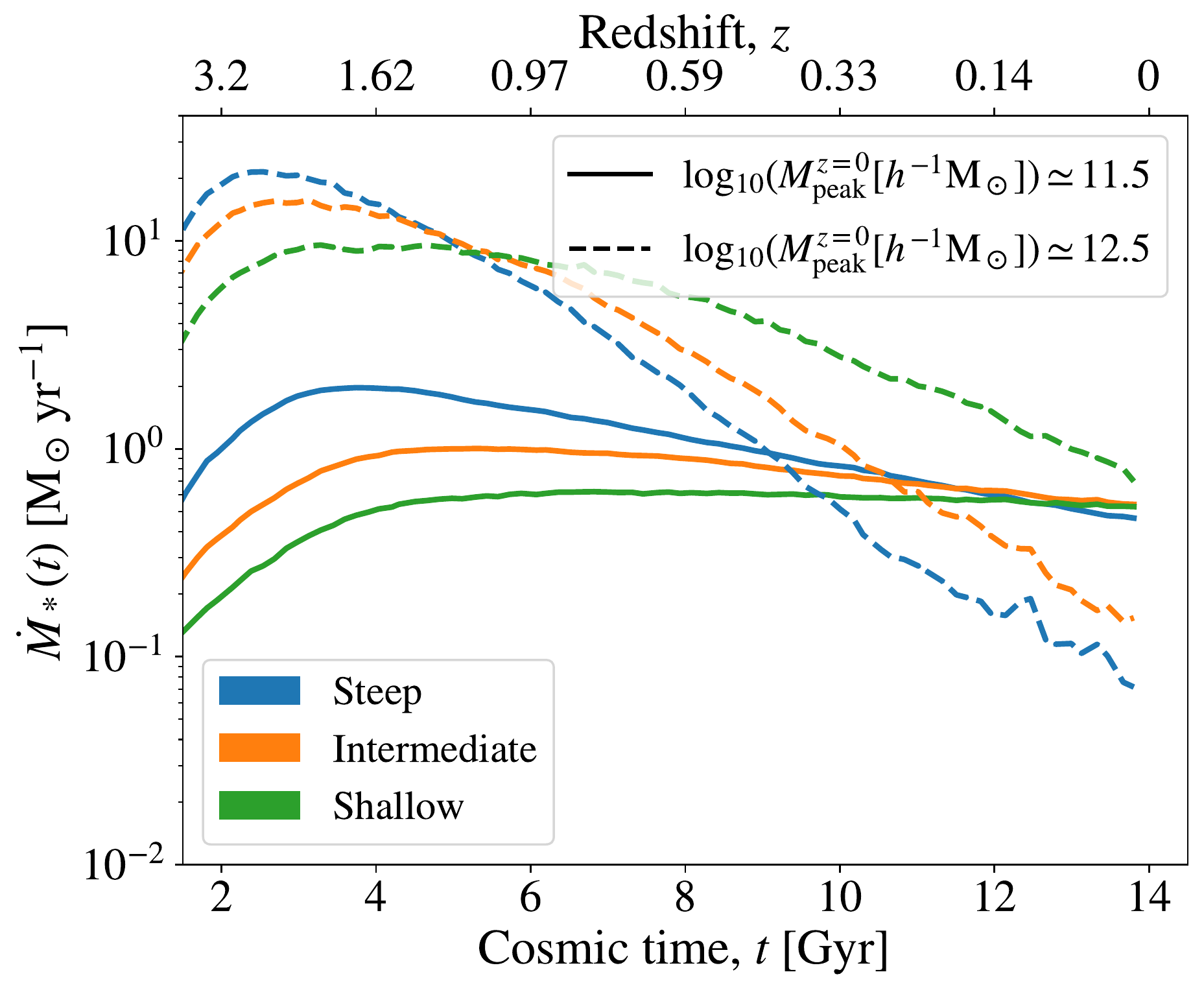}
    \includegraphics[width=0.33\textwidth]{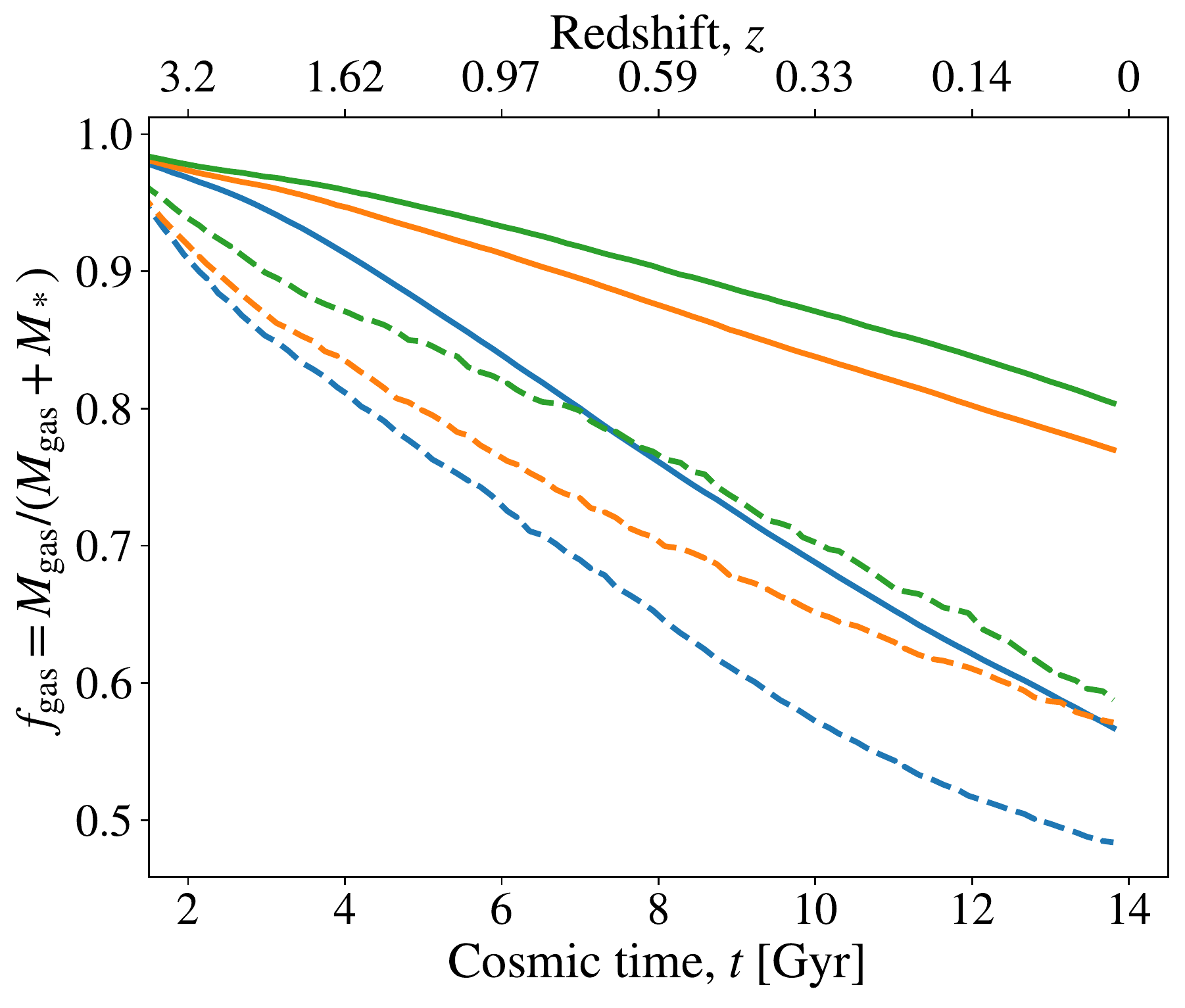}
    \includegraphics[width=0.33\textwidth]{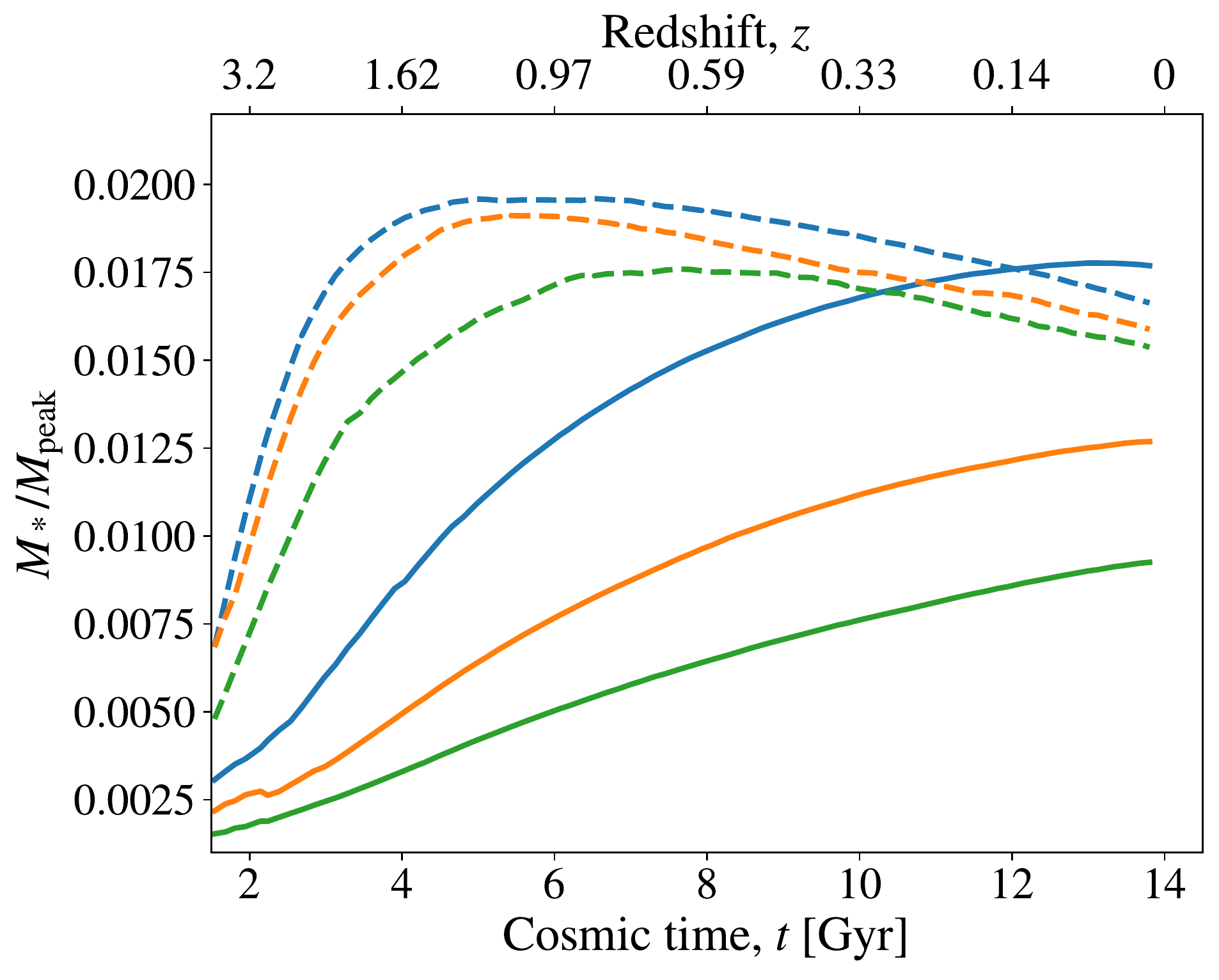}
    
    \begin{tabular}{c}
        \textbf{\Large Formation time (\tform)} \\
    \end{tabular}
    \hspace*{0.75cm}
    
    \includegraphics[width=0.33\textwidth]{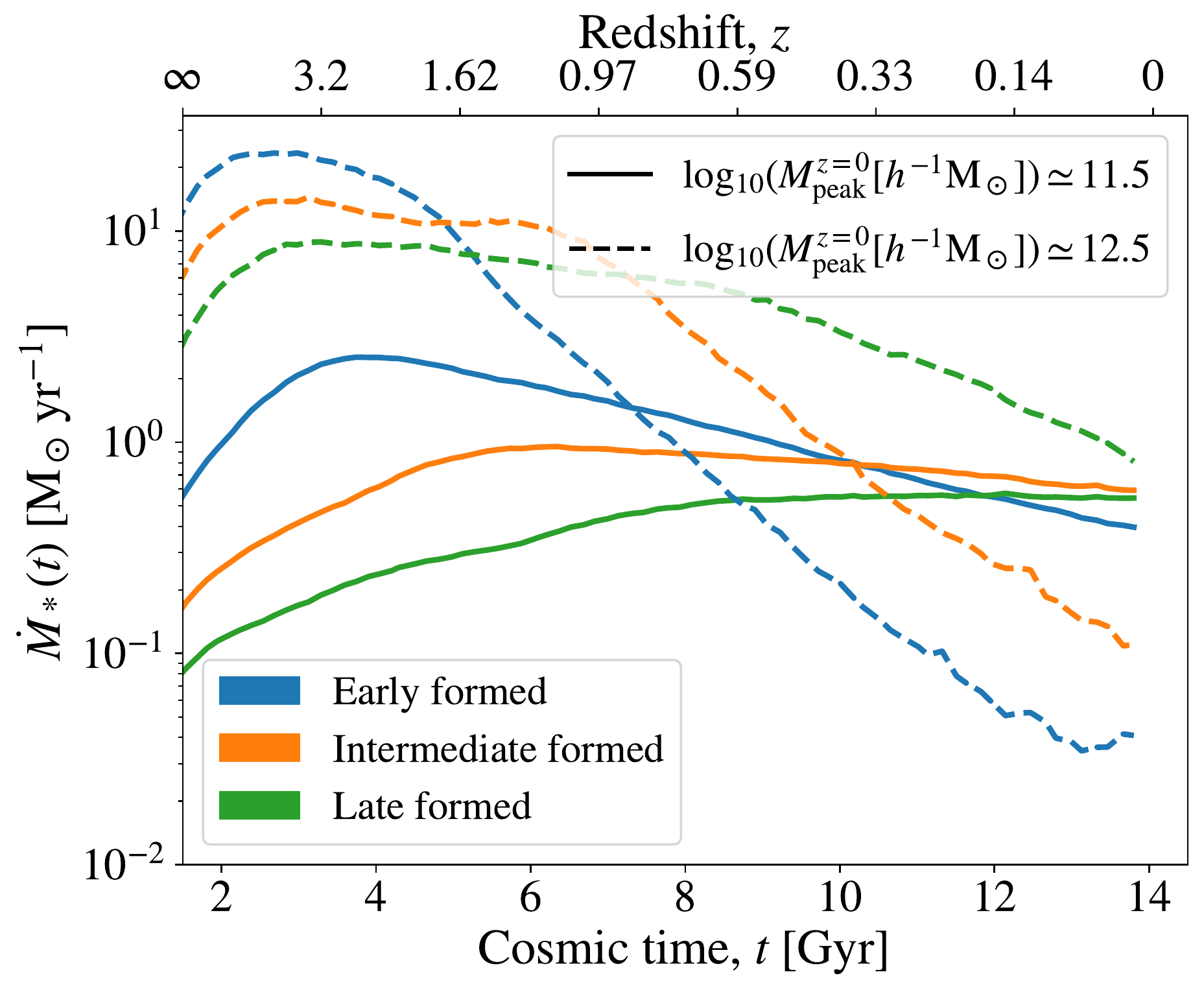}
    \includegraphics[width=0.33\textwidth]{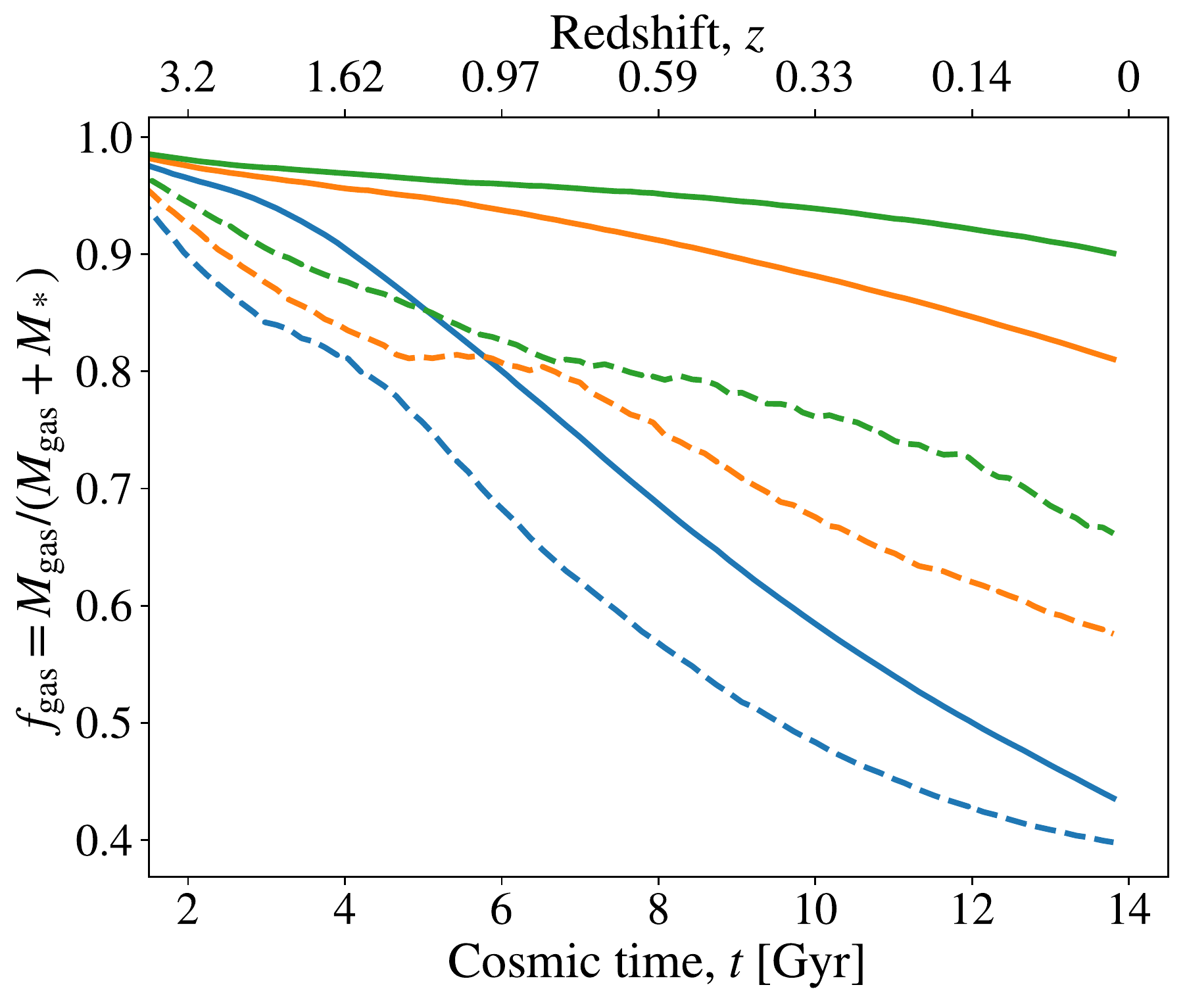}
    \includegraphics[width=0.33\textwidth]{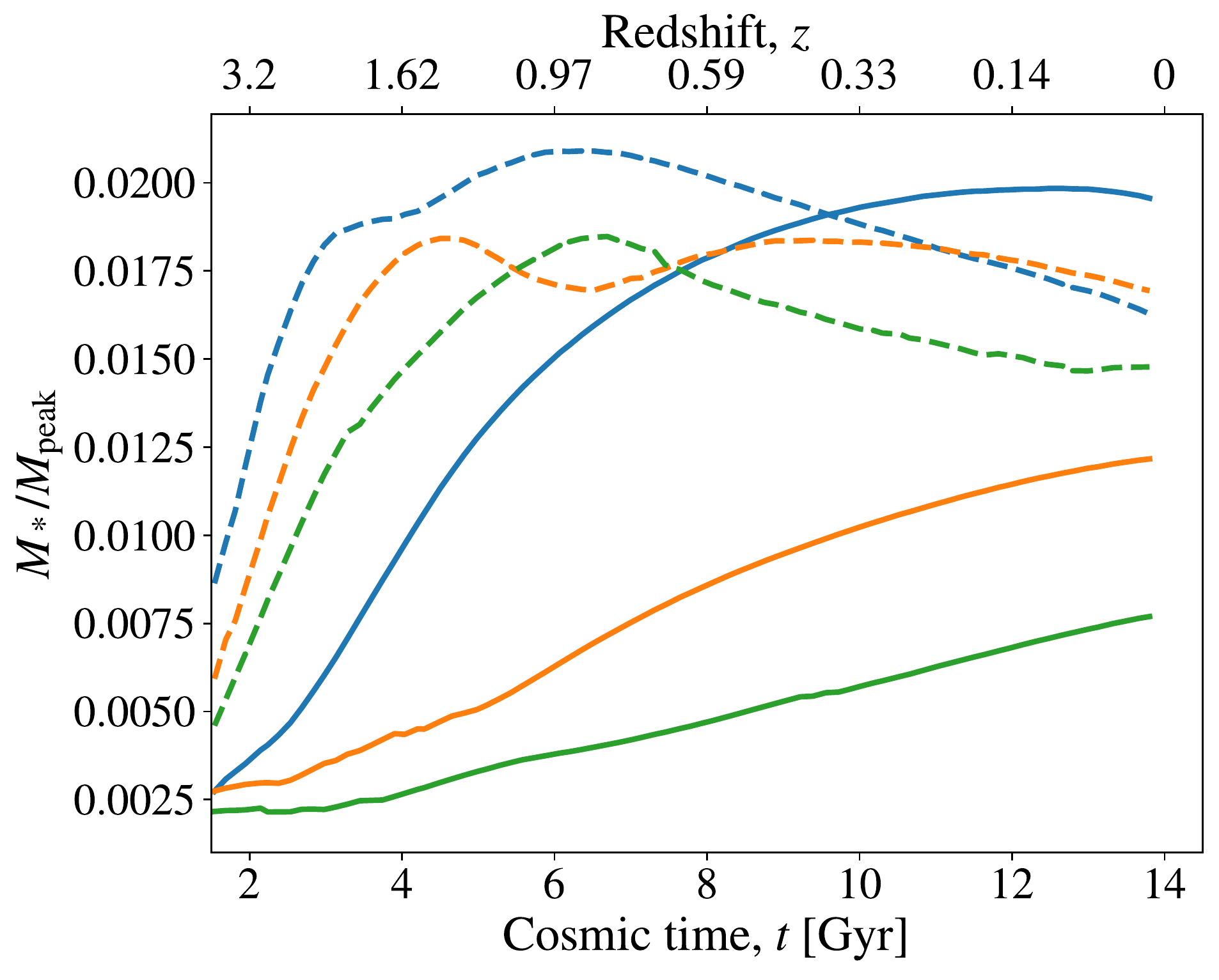}
    
    \caption{Redshift evolution of star formation rate (SFR), gas fraction, and stellar-to-halo mass ratio for \TNG galaxies selected according to the sMAH slope (\slope, top panels) and formation time (\tform, bottom panels). The same format as in the central and right panels of Fig.~\ref{fig:model_sMAH_cartoon} is adopted. The aforementioned properties display monotonic and non-monotonic dependencies on \slope and \tform, respectively, supporting the notion that \slope is more fundamentally connected to the properties of the hosted galaxies than \tform.}
    \label{fig:model_sMAH_tform}
\end{figure*}

\section{Halo mass accretion history}
\label{sec:model}

As mentioned previously, the MAH of DM haloes is the fundamental property regulating the formation and evolution of galaxies in multiple galaxy formation models. However, 
this connection has been proven difficult to establish on the basis of simple features \citep{ChavesMontero2016, Campbell2018_sham, MonteroDorta2020B, Hadzhiyska2021_sham,Favole2021}, which are in most cases also affected by transient effects or subhalo identification and tracking issues \citep{knebe2011_comparison, Knebe2013_comparison, Behroozi2013_grav, ChavesMontero2016, Wang2020_concentration}. In this section, we introduce a new halo assembly feature that aims to alleviate the shortcomings associated with some of the previously used properties.

In the left panel of Fig.~\ref{fig:model_sMAH_cartoon}, the {\it{specific}} mass accretion history, sMAH~$\equiv \Mpeak^{-1} (\mathrm{d}\Mpeak/\mathrm{d}t)$, is displayed for a randomly selected \TNG galaxy. We normalise the accretion history of a halo by its mass to capture the relative growth of haloes independently of their current mass; note that the inverse of the sMAH is a measure of the time it takes a halo to assemble its mass given its current accretion rate. The blue solid line shows the sMAH predicted by \TNG, while the orange dashed line represents a best-fitting model computed by performing a linear fit to the sMAH as a function of cosmic time in logarithmic space. This best-fitting model, as Fig.~\ref{fig:model_sMAH_cartoon} demonstrates, captures the redshift evolution of the data precisely. Overall, we find that \TNG predicts an approximate power-law evolution for most haloes independently of their mass, which strongly supports the use of the slope of the best-fitting power law, which we denote as $\slope$, as a means of characterising the time evolution of sMAHs. Note that this power-law redshift dependence is also predicted by gravity-only cosmological simulations \citep{Behroozi2015}.

In the central and right panels of Fig.~\ref{fig:model_sMAH_cartoon}, the average sMAHs and halo mass histories of a representative group of \TNG galaxies are presented. Solid and dashed lines indicate results obtained for galaxies hosted by $\log_{10}(\Mpeakz [h^{-1}{\rm M_\odot}])=11.5$ and 12.5 haloes, respectively, while
the colour pattern represents the haloes' sMAH slope, \slope. Blue, orange, and green colours correspond to galaxies in haloes with small, intermediate, and large \slope (each \slope subset containing approximately 1/3 of the total sample), respectively. We find that galaxies in haloes of increasingly steeper (shallower) \slope present more (less) accretion at early (late) times. In other words, haloes with steeper \slope accrete most of their mass at earlier times as compared to those with shallower slopes. Consequently, it is natural to expect correlation between \slope and halo formation time (\tform). In fact, we have checked that the Spearman rank-order correlation coefficient between these quantities is typically larger than 0.7.

Motivated by these findings, in Fig.~\ref{fig:model_sMAH_tform} we present the average redshift evolution of the SFR, gas fraction, and stellar-to-halo mass ratio for \TNG galaxies selected according to \slope (top row) and \tform (bottom row). We use the same format and colour code as in the central and right panels of Fig.~\ref{fig:model_sMAH_cartoon}. The solid lines show that the dependence of the redshift evolution of these properties on \slope is similar to that of \tform when galaxies hosted by $\log_{10}(\Mpeakz [h^{-1}{\rm M_\odot}])=11.5$  haloes are considered. On the contrary, the average dependence of galaxy properties on \tform is non-monotonic at $z<1$ for galaxies hosted by $\log_{10}(\Mpeakz [h^{-1}{\rm M_\odot}])=12.5$ haloes, while the dependence of these properties on \slope is monotonic at all times. This result indicates that the sMAH slope presents a more fundamental connection with the properties of the hosted galaxies than formation time.

Fig.~\ref{fig:model_sMAH_tform} also shows that at fixed halo mass galaxies with increasingly steeper \slope reach the peak of their SFHs at higher redshift, present larger stellar-to-halo mass ratios, and lose their gas faster, thus becoming quenched earlier. It is noteworthy that the dependence of these galaxy properties on \slope is similar to the impact that the so-called {\it{downsizing}} have on them. Namely, galaxies hosted by more massive haloes reach the peak of their SFH earlier, present a larger stellar mass, and quench faster \citep[e.g.,][]{cowie1996_NewInsightGalaxy, conroy2009_CONNECTINGGALAXIESHALOS, fontanot2009_ManyManifestationsdownsizing, dave2016_MUFASAGalaxyformation, chaves_hearin2020_sbu1}. Therefore, the slope of the sMAH seems to capture the downsizing signal at fixed halo mass, which can explain residual correlation between halo and galaxy assembly that halo mass is unable to capture. Altogether, these findings motivate the use of \slope to investigate the dependence of galaxy clustering on the assembly history of haloes, an aspect that will be addressed in the following section.

\begin{figure*}
  \centering
  \includegraphics[width=\textwidth]{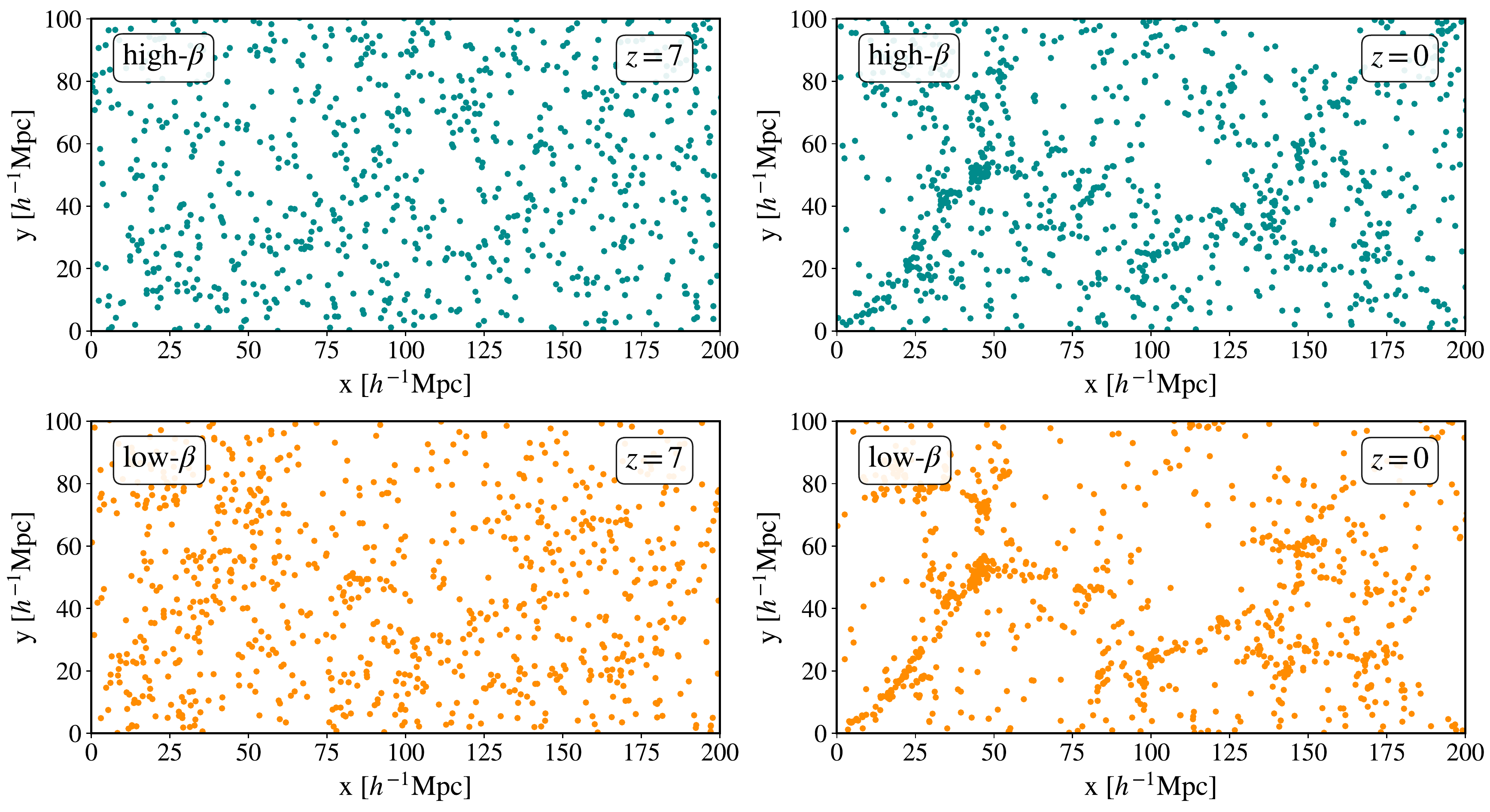}

\caption{Spatial distribution of TNG300 galaxies hosted by high-\slope (upper panels) and low-\slope (lower panels) haloes in a slice of 200x100x20 $h^{-3}$Mpc$^{3}$. For the sake of visibility, only a small randomly taken subset of galaxies in haloes of masses in the range $\log_{10}(\Mpeakz [h^{-1} {\rm M}_{\odot}]) = 11.5 - 11.7 $ are included. From left to right, their evolution is mapped from redshifts $z=7$ and 0. Galaxies in haloes with lower \slope appear more clustered than those in high-\slope haloes, particularly at $z=0$.}
\label{fig:distrib}
\end{figure*}

\section{The dependence of galaxy clustering on halo mass accretion history}
\label{sec:results}

At fixed halo mass, the clustering of DM haloes depends, among other properties, on their assembly histories, an effect commonly called halo assembly bias (e.g., \citealt{gao2005,wechsler2006,2018Salcedo,SatoPolito2019}). This particular secondary dependence is also expected to manifest itself on the clustering of the galaxy population, even though a definite observational proof of this connection is yet to be established (see, e.g., \citealt{Lin2016,MonteroDorta2017B,Niemiec2018,Salcedo2020,Obuljen2020}). As mentioned in the Introduction, the term galaxy assembly bias designates throughout this work the dependence of galaxy clustering on the MAH of their hosting haloes at fixed halo mass (thus excluding other potential contributions such as those arising from, e.g., spin bias, see \citealt{Tucci2020}).

In this section, we measure the halo assembly bias signal ``transmitted" to the galaxy population. To this end, we set a minimum stellar mass $\log_{10} (M_*(t)[h^{-1} {\rm M_\odot}]) > 7$ so that only subhaloes which are effectively hosting galaxies are selected. Note that the initial gas-cell resolution of TNG300 is $\log_{10} (M_{\rm gas}(t)[ h^{-1}\Msun]) \simeq 6.88$, which implies that we are also selecting ``unresolved" galaxies. However, we want to collect all progenitors of present-day galaxies, and a more restrictive threshold would result in severe incompleteness at high redshift. Importantly, we are not interested in splitting the galaxy sample by stellar mass (or any other galaxy property), so we expect this caveat to have a minor impact on our results. By construction, these selection criteria result in a galaxy assembly bias signal necessary similar to the halo assembly bias signal of their host haloes.

In Fig.~\ref{fig:distrib}, we illustrate the impact of assembly bias on galaxy clustering. Cyan and orange dots represent a small randomly taken subset of galaxies satisfying the aforementioned selection criteria and hosted by haloes in the high and low tail of the \slope distribution, respectively, at both $z=7$ (left panels) and 0 (right panels). Recall that low-$\beta$ values typically correspond to older haloes (since they more rapidly accrete mass at early times), which explains why the low-\slope objects in Figure~\ref{fig:distrib} appear more clustered than their high-\slope counterparts, particularly at $z=0$.  

A consequence of the results presented in Section~\ref{sec:model} is that the sMAH slope seems ideally suited for the study of assembly bias, since: 1) it shows a strong correlation with multiple halo and galaxy properties, 2) it is robust against subhalo identification and tracking issues as well as rapid changes in halo properties  due to mergers, and 3) it enables easily connecting subhaloes across redshift (as a consequence of the negligible evolution of \slope with redshift).

\begin{figure*}
    \centering
    \includegraphics[scale=0.49]{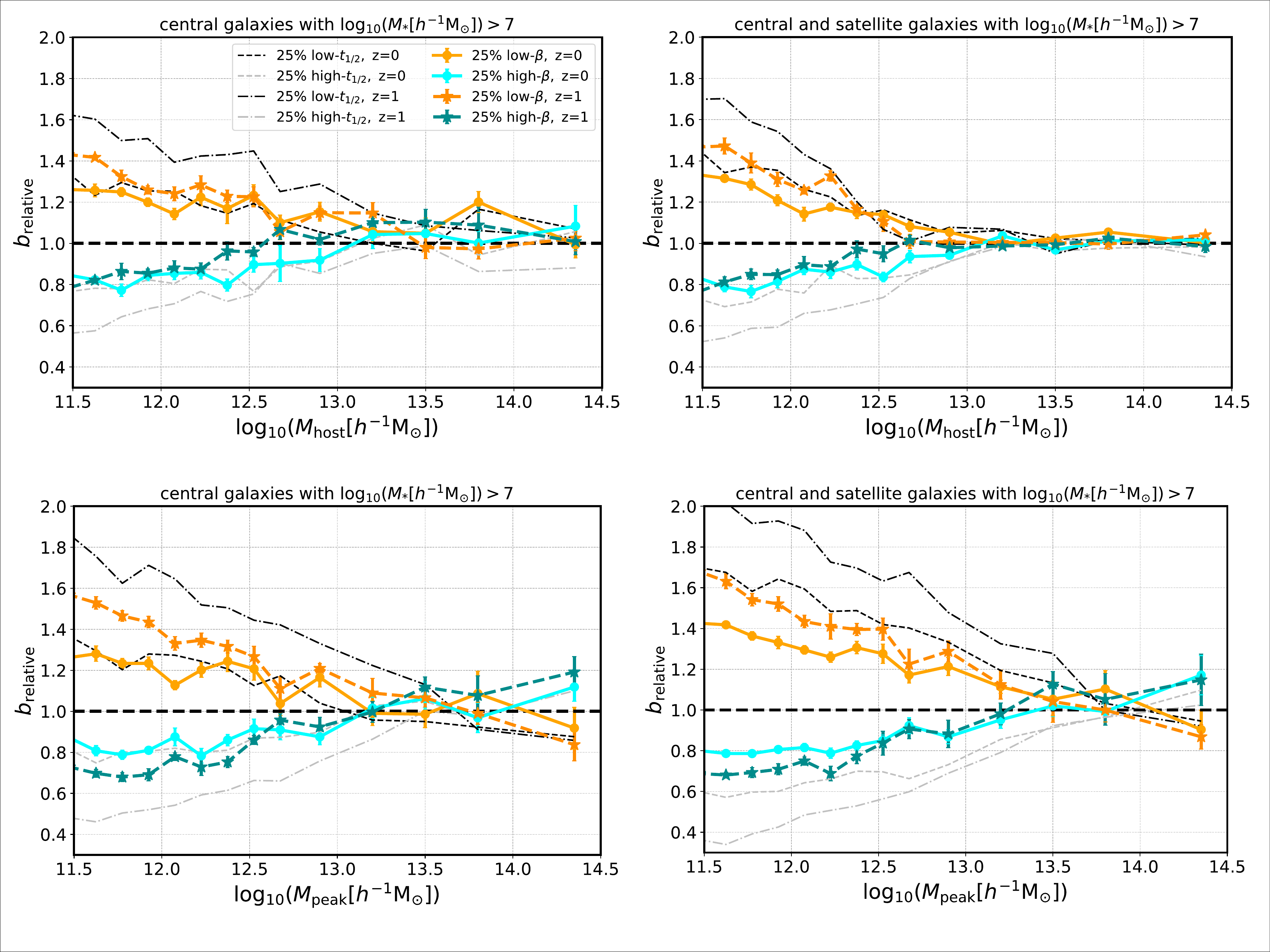}
    \caption{The secondary dependence of galaxy bias on the shape of the sMAH from TNG300. 
    In each panel, the relative bias for quartile subsets
    selected on the basis of \slope and the entire population at the corresponding mass bin is shown for $z=0$ and $z=1$. For reference, the secondary dependence on formation time, \tform, is also plotted. 
    The upper panels display the relative bias as a function of the mass of the host halo at the corresponding redshift, $M_{\rm host}(z)$, whereas the peak mass of the host, \Mpeakz, is chosen as the primary halo property in the lower panels. In the right-hand panels, both centrals and satellites are included, whereas only centrals are considered in the left-hand panels. All relative bias measurements are averaged on scales $5< r[h^{-1}{\rm Mpc}]<15$. Error bars represent the uncertainties on the measurements obtained from a jackknife technique (see text).}
    \label{fig:bias_mass}
\end{figure*}

\begin{figure*}
    \centering
    \includegraphics[scale=0.6]{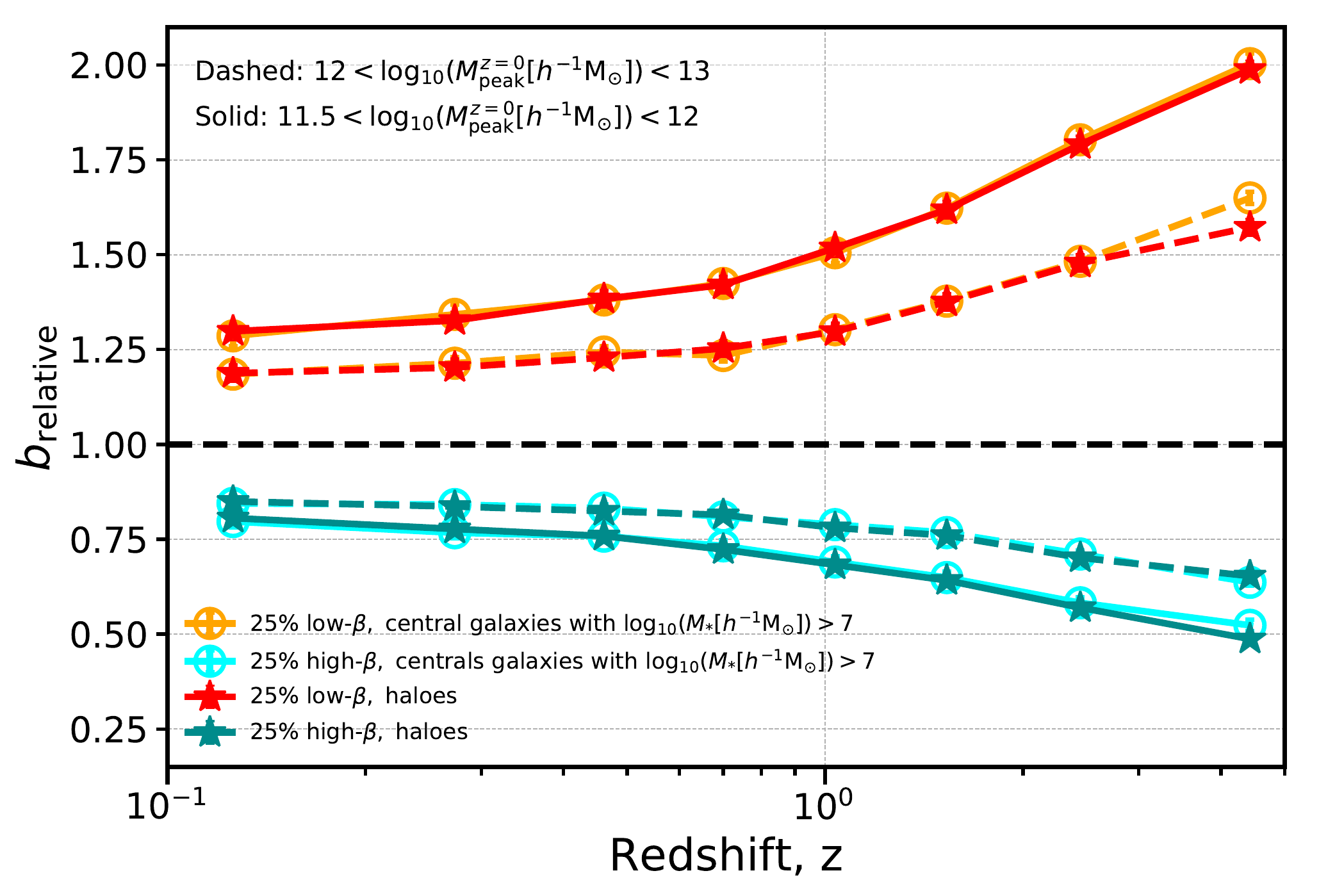}
    \caption{The redshift evolution of the secondary dependence of galaxy clustering on \slope, integrating over two different \Mpeakz mass ranges. Error bars represent the uncertainties on the measurements obtained from a jackknife technique (see text).}
    \label{fig:bias_z}
\end{figure*}

In order to analyse in more detail the assembly bias signal produced by the \slope parameter at fixed halo mass, we follow a common procedure and measure the {\it{relative bias}}, $b_{\rm relative}$, between conveniently chosen subsets of galaxies \citep{2018Salcedo}. It is noteworthy that TNG300 is one of the largest hydrodynamical boxes available to the community and one of the first where statistically significant measurements of clustering can be performed (for certain scales and mass ranges, see, e.g., \citealt{MonteroDorta2020B, MonteroDorta2020C}). In order to compute the relative bias for different subsets of galaxies, we measure the 3D two-point correlation function, $\xi(r)$, using the Landay-Szalay estimator \citep{Landy1993}. Following \cite{MonteroDorta2020B}, the relative bias between galaxy subsets is measured on the basis of the cross-correlation between quartiles and the entire halo sample, which maximises the signal-to-noise of the measurement (this is important given the still relatively small size of TNG300). For a given halo-mass bin $M_{i}$ and subset $\mathcal{S}$, the relative bias can be measured as: 

\begin{equation}
   b_{\rm relative}(r,\mathcal{S}|M_i) = \frac{\xi_{[\mathcal{S},{\rm all}]}(r)}{\xi_{[M_{i},{\rm all}]}(r)},
   \label{eq:bias2}
\end{equation}
 
\noindent where $\xi_{[S,{\rm all}]}$ is the cross-correlation between all objects in the subset and all objects in the sample, and $\xi_{[ M_i,{\rm all}]}$ is the cross-correlation between all objects in the halo-mass bin and the entire sample as well. The computation of errors is based on a standard jackknife technique, where the TNG300 box is divided in 8 sub-boxes ($ L_{{\rm sub-box}} = L_{{\rm box}}/2 = 102.5$ $h^{-1}$Mpc). As mentioned in the Introduction, the relative bias is averaged over scales 5-15 $h^{-1}$Mpc.

Eq.~\ref{eq:bias2} is employed in Fig.~\ref{fig:bias_mass}, which displays the secondary dependence of galaxy clustering on \slope as a function of the primary property, i.e., halo mass. Here, the analysis is restricted to the conservative mass range $\log_{10}( M_{\rm halo}[h^{-1}{\rm M_\odot}]) > 11.5$ in order to avoid resolution problems for lower mass haloes (and their corresponding central galaxies, see \citealt{MonteroDorta2020B}). In all panels, the relative biases for quartile subsets of galaxies defined in terms of \slope are shown as a function of halo mass (25$\%$ higher/smaller values in cyan/orange tones), for two different redshift snapshots ($z=0$ and 1). Upper panels adopt the mass of the halo at the corresponding redshift, $M_{\rm host}$, as a halo-mass definition, whereas lower panels employ \Mpeakz (note that $M_{\rm host}$ is redshift-dependent). The right-hand panels show results for the combined sample (centrals and satellites) while the measurement is restricted to central galaxies on the left-hand side. Importantly, central and satellite galaxies are selected at the corresponding redshift of the measurement. 

The evolution of galaxy assembly bias is measured here for the progenitors of present-time galaxies. This {\it{evolution of galaxy assembly bias along the merger tree}} is conceptually different from measuring the effect for the entire galaxy population at each snapshot (or halo population, see, e.g., \citealt{Gao2007}). We maintain the same nomenclature for simplicity, but it is important to bear in mind the differences with respect to previous approaches.

We begin by addressing the central-galaxy sample. Independently of the redshift snapshot considered, galaxies that live in haloes with low \slope (steep slope) are more tightly clustered than their slowly evolving counterparts. As expected, this signal seems to vanish towards the high-mass end (although the size of TNG300 imposes an obvious statistical limitation here). Using peak mass \Mpeakz or the instantaneous mass $M_{\rm host}$ has an impact on the amplitude of the \slope dependence. The signal is stronger when the measurement is performed as a function of \Mpeakz, particularly towards the high-mass end. 

Overplotted in Fig.~\ref{fig:bias_mass} is the secondary dependence on \tform for the two redshift slices considered (in black lines; this is the ``standard" proxy for age used in the literature in the context of halo assembly bias, see, e.g., \citealt{gao2005,2018Salcedo,SatoPolito2019}). As expected, since \slope correlates with \tform, both trends have very similar shapes. Interestingly, however, the signals for \slope are significantly attenuated with respect to the \tform results, particularly at high redshift.  

The inclusion of satellite galaxies that pass our minimum mass threshold (right-hand panels) has little impact on the trends qualitatively, as it could be expected given the range of scales employed in the bias measurement. However, some small differences are noticeable. In particular, satellites tend to attenuate the signal towards the high-mass end (particularly for $M_{\rm host}$), while accentuating it for lower mass haloes (see, e.g., the bottom-right panel of Fig.~\ref{fig:bias_mass}). The fact that satellites increase the signal in some mass ranges is interesting since they are usually regarded as contaminants of the assembly bias signal. 

Fig.~\ref{fig:bias_mass} shows significant redshift evolution of the secondary \slope dependence. Note that the underlying halo assembly bias signal measured from $N$-body numerical simulations (based on formation time) is known to decrease with increasing redshift at fixed (instantaneous) halo mass, when the entire halo population at each snapshot is considered (see, e.g., \citealt{Gao2007}). This evolution is consistent with the signal depending on mass only through the peak height of the fluctuations, $\nu = \delta_c(z)/\sigma(M_{\rm halo},z)$, where $\delta_c(z)$ is the redshift-dependent background density contrast and $\sigma(M_{\rm halo},z)$ is the variance of the linear overdensity field on a sphere containing a mass of  $M_{\rm halo}$ at redshift $z$. The progenitor evolution presented in Fig.~\ref{fig:bias_mass} displays the opposite trend, which can be understood from the right-hand panel of Fig.~\ref{fig:model_sMAH_cartoon}. In essence, haloes of a given mass at $z = 0$ that experienced a steeper MAH (low value of $\beta$) were typically more massive at $z = 1$ than those that ``arrived" at the same $z=0$ mass via a shallow MAHs (high value of \slope). More massive haloes imply, on average, higher large-scale biases \citep{ShethTormen2002}, which explains the measured trends. 

When $M_{\rm host}(z)$ is used as the binning property, the signal still increases with redshift, but not so rapidly as in the \Mpeakz case. The reason is that the progenitor trend explained above is slightly compensated by the fact that the  
assembly bias signal for a fixed halo mass decreases with redshift \citep{Gao2007}. This effect is noticeable at the high-mass end, where the first ``crossing" of the signal (the mass at which the ratio converges to 1)  tends to occur at lower masses for the $z=1$ measurement.
 
At a reference mass range $\log_{10}( \Mpeakz [h^{-1}{\rm M_\odot}]) = 11.5 - 12$, the amplitude of the signal increases by $\sim$ 15-25 $\%$ for each individual quartile subset between $z=0$ and 1. The redshift evolution of galaxy assembly bias for central galaxies is explored in more detail in Fig.~\ref{fig:bias_z}, where the relative bias is averaged over two mass ranges: $\log_{10}( \Mpeakz [h^{-1}{\rm M_\odot}]) = 11.5 - 12$ and 12 - 13. Fig.~\ref{fig:bias_z} shows that the \slope dependence of clustering increases at a moderate pace for both mass bins up to $z\sim0.6$. After $z\sim0.6-1$, however, the signal starts amplifying significantly, reaching a relative value (between quartile subsets) of $\sim 2/0.5 = 4$ at redshift $z = 4.5$ for the less massive bin. At $z=1$, where several cosmological surveys will overlap in the near future, and for haloes of peak mass $\log_{10}( \Mpeakz [h^{-1}{\rm M_\odot}]) = 11.5 - 12$, the signal reaches a factor of approximately 2 (i.e., $\sim$ 1.5/0.75). Also plotted in Fig.~\ref{fig:bias_z} is the halo assembly bias trend, which corresponds to the case where no stellar mass threshold is imposed on subhaloes. As expected, only at very high redshifts does this condition impact the galaxy assembly bias signal, as incompleteness becomes more significant.

\section{Discussion \& Conclusions}
\label{sec:discussion}

\begin{figure*}
    \centering
    \includegraphics[width=0.475\textwidth]{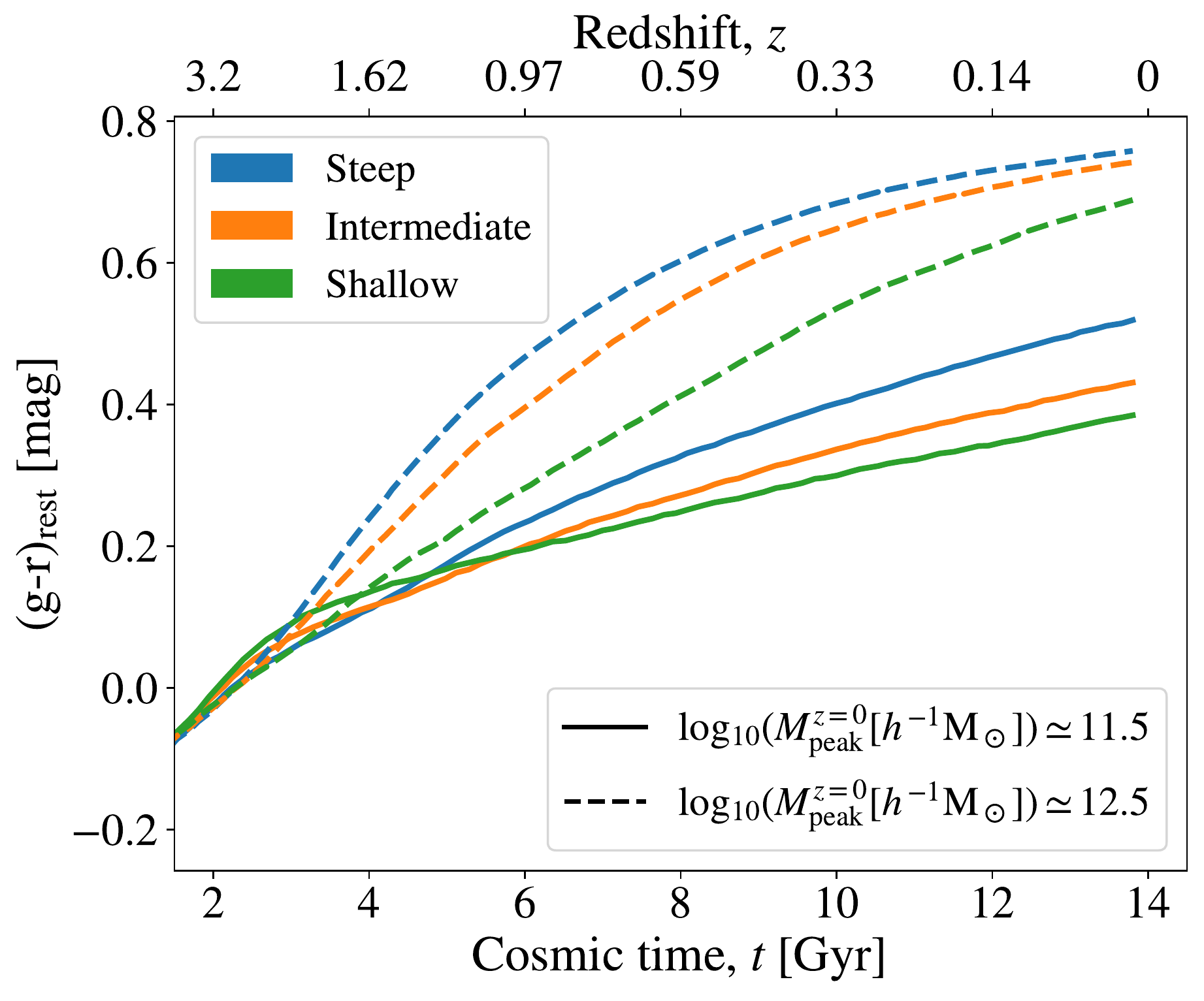}
    \includegraphics[width=0.475\textwidth]{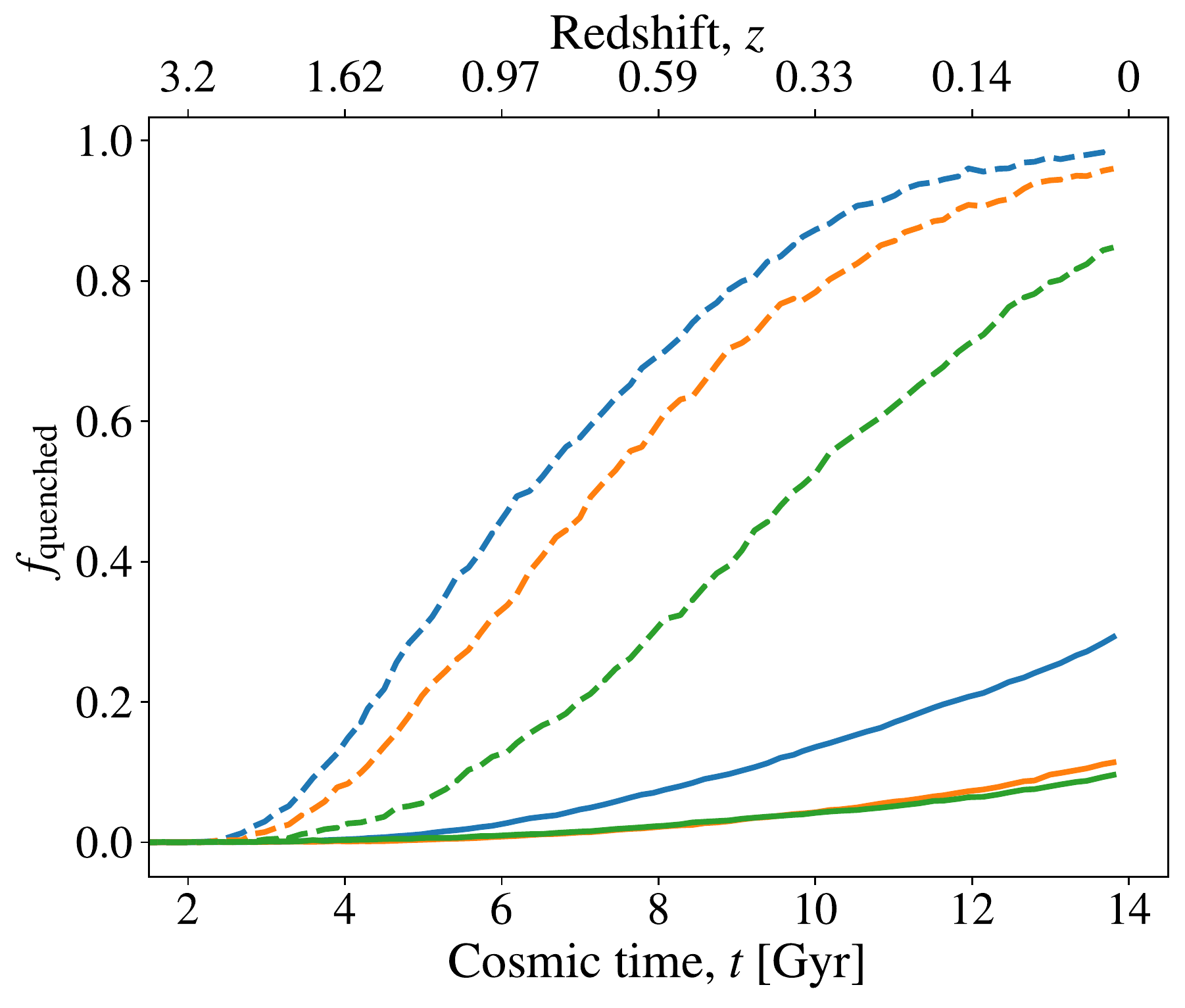}
    
    \caption{Dependence of galaxy colours and quenched fraction on \slope. Solid and dashed lines indicate results for galaxies hosted by haloes of mass $\log_{10}( \Mpeakz [h^{-1}{\rm M_\odot}]) = 11.5$ and 12.5, respectively, while blue, orange, and green colours do so for galaxies with steep, intermediate, and shallow \slope. As we can see, galaxies with increasingly steeper \slope present redder colours and quench at earlier times.}
    \label{fig:model_sMAH_prop}
\end{figure*}

We have shown that the slope of the specific mass accretion history (sMAH), \slope, is advantageous as a means of describing halo evolution in that it displays monotonic and smooth relations with important galaxy properties, including SFR, gas fraction, and stellar-to-halo mass ratio. Our results also demonstrate that \slope provides a more direct link to the galaxy formation process, represented here by the aforementioned properties, than the commonly used \tform parameter.

These advantages can be exploited in order to generate high-fidelity mock catalogues for current and future cosmological surveys in a more efficient way. In this context, Fig.~\ref{fig:model_sMAH_prop} shows, in the same format as in Fig.~\ref{fig:model_sMAH_tform}, the dependence of two galaxy observables on \slope. The left-hand panel demonstrates that the slope of the sMAH provides a ``clean" link to the rest-frame (g-r) galaxy colour, at least up to $z\sim1.6$. Galaxies that live in haloes with steeper sMAH present redder colours. Consequently, \slope also correlates very well with the {\it{quenched fraction}}, i.e., the fraction of quenched (non star-forming) galaxies of the total galaxy population, as the right-hand panel of Fig.~\ref{fig:model_sMAH_prop} indicates. Here, galaxies are considered quenched when their specific SFR is lower than $10^{-11}\mathrm{yr}^{-1}$. Galaxies that live in haloes with steeper \slope also quench at earlier times. One particular way in which \slope can be helpful is as a secondary halo property in
modified versions of halo-galaxy connection models such as sub-halo abundance matching (SHAM, e.g., \citealt{Hearin2013,Hadzhiyska2021_sham,Favole2021}) and halo occupation distributions (HODs, e.g., \citealt{Hearin2016,Xu2021,Hadzhiyska2020,Hadzhiyska2021}).

We have also measured {\it{galaxy assembly bias}}, understood here simply as the dependence of galaxy bias on \slope at fixed halo mass. This definition implies a direct manifestation of halo assembly bias, thus excluding other potential dependencies arising from other secondary halo properties (e.g. spin). The theoretical approach that we follow here differs from previous analyses in that we determine the redshift evolution of assembly bias for the progenitors of present-day galaxies. We show that the signal increases significantly at high redshift (contrary to the evolution measured for halo assembly bias in DM-only simulations for independent snapshots and the entire population). This increase in signal at high redshift is due to progenitors acquiring mass faster for steeper-sMAH (typically older) haloes.

A consequence of our findings is that, at fixed mass, individual haloes that are more clustered at $z=0$ are also more clustered at high redshift; i.e., halo/galaxy assembly bias is imprinted in the distribution of haloes/galaxies from early on. This is somewhat expected given that haloes do not move around their initial positions more than a few Mpc during their lifespan. This work is, in fact, the first to our knowledge to analyse the time dependence of assembly bias for individual objects.

Most current and next-generation cosmological surveys such as the Dark Energy Spectroscopic Instrument (DESI\footnote{https://www.desi.lbl.gov}), Euclid\footnote{\url{https://www.euclid-ec.org}}, the Javalambre Physics of the Accelerated Universe Astrophysical Survey (J-PAS\footnote{\url{http://www.j-pas.org}}), or the Vera C. Rubin Observatory (also/formerly known as LSST\footnote{https://www.lsst.org}) will focus on redshift ranges that span anywhere from $z\sim 1$ to, in some cases, $z\sim 3$. Several of these projects will also target emission-line galaxies, which tend to live in lower-mass haloes for which the assembly bias signal is large. Our results regarding the evolution of assembly bias might have implications for future high-redshift cosmological measurements that are worth exploring. Further work will be also devoted to evaluating potential applications in the context of observational probes of assembly bias. 

We have shown that different proxies for the shape of the MAH at fixed halo mass produce slightly different galaxy assembly bias signals. This is not surprising since the underlying halo assembly bias effect is known to be very sensitive to this choice (e.g., halo concentration, \citealt{wechsler2006}; different definitions of halo age, \citealt{Chue2018}).
In this context, \cite{Borzyszkowski2017} and \cite{Musso2018} describe a mechanism by which halo assembly bias, in its multiple manifestations, would be the consequence of the truncation of halo accretion history preferentially in some cosmic-web environments as compared to others (i.e., filaments vs. nodes). In a preliminary analysis, we have checked that the \slope parameter appears to display little or no correlation with the anisotropy parameter $\alpha$ (which can be used to define cosmic-web environments, see, e.g., \citealt{Ramakrishnan2019}). This connection between \slope, the cosmic-web environment, and assembly bias can be further explored. In principle, \slope can be used to identify subpopulations of haloes with a truncated accretion history and study the effect that removing these haloes has on the assembly bias trends.

\section*{Acknowledgments}

We thank Andrew Hearin and Beatriz Tucci for their useful comments. MCA acknowledges financial support from the Austrian National Science Foundation through FWF stand-alone grant P31154-N27. GF acknowledges financial support from the SNF 175751 “Cosmology with 3D Maps of the Universe” research grant. 
\section*{Data availability}
The simulation data underlying this article are publicly
available at the TNG website. The data results arising from this work will be shared on reasonable request to the corresponding authors.

\bibliographystyle{mnras}
\bibliography{biblio}

\begin{thebibliography}{}
\makeatletter
\relax
\def\mn@urlcharsother{\let\do\@makeother \do\$\do\&\do\#\do\^\do\_\do\%\do\~}
\def\mn@doi{\begingroup\mn@urlcharsother \@ifnextchar [ {\mn@doi@}
  {\mn@doi@[]}}
\def\mn@doi@[#1]#2{\def\@tempa{#1}\ifx\@tempa\@empty \href
  {http://dx.doi.org/#2} {doi:#2}\else \href {http://dx.doi.org/#2} {#1}\fi
  \endgroup}
\def\mn@eprint#1#2{\mn@eprint@#1:#2::\@nil}
\def\mn@eprint@arXiv#1{\href {http://arxiv.org/abs/#1} {{\tt arXiv:#1}}}
\def\mn@eprint@dblp#1{\href {http://dblp.uni-trier.de/rec/bibtex/#1.xml}
  {dblp:#1}}
\def\mn@eprint@#1:#2:#3:#4\@nil{\def\@tempa {#1}\def\@tempb {#2}\def\@tempc
  {#3}\ifx \@tempc \@empty \let \@tempc \@tempb \let \@tempb \@tempa \fi \ifx
  \@tempb \@empty \def\@tempb {arXiv}\fi \@ifundefined
  {mn@eprint@\@tempb}{\@tempb:\@tempc}{\expandafter \expandafter \csname
  mn@eprint@\@tempb\endcsname \expandafter{\@tempc}}}

\bibitem[\protect\citeauthoryear{{Angulo}, {Baugh}  \& {Lacey}}{{Angulo}
  et~al.}{2008}]{Angulo2008}
{Angulo} R.~E.,  {Baugh} C.~M.,   {Lacey} C.~G.,  2008, \mn@doi [\mnras]
  {10.1111/j.1365-2966.2008.13304.x}, \href
  {http://adsabs.harvard.edu/abs/2008MNRAS.387..921A} {387, 921}

\bibitem[\protect\citeauthoryear{{Artale}, {Zehavi}, {Contreras}  \&
  {Norberg}}{{Artale} et~al.}{2018}]{Artale2018}
{Artale} M.~C.,  {Zehavi} I.,  {Contreras} S.,   {Norberg} P.,  2018, \mn@doi
  [\mnras] {10.1093/mnras/sty2110}, \href
  {https://ui.adsabs.harvard.edu/abs/2018MNRAS.480.3978A} {480, 3978}

\bibitem[\protect\citeauthoryear{{Avila-Reese}, {Firmani}  \&
  {Hern{\'a}ndez}}{{Avila-Reese} et~al.}{1998}]{avila1998_sam}
{Avila-Reese} V.,  {Firmani} C.,   {Hern{\'a}ndez} X.,  1998, \mn@doi [\apj]
  {10.1086/306136}, \href
  {https://ui.adsabs.harvard.edu/abs/1998ApJ...505...37A} {505, 37}

\bibitem[\protect\citeauthoryear{{Baugh}, {Cole}  \& {Frenk}}{{Baugh}
  et~al.}{1996}]{baugh1996_sam}
{Baugh} C.~M.,  {Cole} S.,   {Frenk} C.~S.,  1996, \mn@doi [\mnras]
  {10.1093/mnras/283.4.1361}, \href
  {https://ui.adsabs.harvard.edu/abs/1996MNRAS.283.1361B} {283, 1361}

\bibitem[\protect\citeauthoryear{{Becker}}{{Becker}}{2015}]{becker2015_eam}
{Becker} M.~R.,  2015, arXiv:1507.03605, \href
  {https://ui.adsabs.harvard.edu/abs/2015arXiv150703605B} {p. arXiv:1507.03605}

\bibitem[\protect\citeauthoryear{{Behroozi} \& {Silk}}{{Behroozi} \&
  {Silk}}{2015}]{Behroozi2015}
{Behroozi} P.~S.,  {Silk} J.,  2015, \mn@doi [\apj]
  {10.1088/0004-637X/799/1/32}, \href
  {https://ui.adsabs.harvard.edu/abs/2015ApJ...799...32B} {799, 32}

\bibitem[\protect\citeauthoryear{{Behroozi}, {Wechsler}, {Wu}, {Busha},
  {Klypin}  \& {Primack}}{{Behroozi} et~al.}{2013}]{Behroozi2013_grav}
{Behroozi} P.~S.,  {Wechsler} R.~H.,  {Wu} H.-Y.,  {Busha} M.~T.,  {Klypin}
  A.~A.,   {Primack} J.~R.,  2013, \mn@doi [\apj] {10.1088/0004-637X/763/1/18},
  \href {https://ui.adsabs.harvard.edu/abs/2013ApJ...763...18B} {763, 18}

\bibitem[\protect\citeauthoryear{{Behroozi}, {Wechsler}, {Hearin}  \&
  {Conroy}}{{Behroozi} et~al.}{2019}]{behroozi2019_um}
{Behroozi} P.,  {Wechsler} R.~H.,  {Hearin} A.~P.,   {Conroy} C.,  2019,
  \mn@doi [\mnras] {10.1093/mnras/stz1182}, \href
  {https://ui.adsabs.harvard.edu/abs/2019MNRAS.488.3143B} {488, 3143}

\bibitem[\protect\citeauthoryear{{Beltz-Mohrmann}, {Berlind}  \&
  {Szewciw}}{{Beltz-Mohrmann} et~al.}{2020}]{Beltz-Mohrmann2020}
{Beltz-Mohrmann} G.~D.,  {Berlind} A.~A.,   {Szewciw} A.~O.,  2020, \mn@doi
  [\mnras] {10.1093/mnras/stz3442}, \href
  {https://ui.adsabs.harvard.edu/abs/2020MNRAS.491.5771B} {491, 5771}

\bibitem[\protect\citeauthoryear{Borzyszkowski, Porciani, Romano-Díaz  \&
  Garaldi}{Borzyszkowski et~al.}{2017}]{Borzyszkowski2017}
Borzyszkowski M.,  Porciani C.,  Romano-Díaz E.,   Garaldi E.,  2017, \mn@doi
  [MNRAS] {10.1093/mnras/stx873}, 469, 594–611

\bibitem[\protect\citeauthoryear{{Bose}, {Eisenstein}, {Hernquist},
  {Pillepich}, {Nelson}, {Marinacci}, {Springel}  \& {Vogelsberger}}{{Bose}
  et~al.}{2019}]{Bose2019}
{Bose} S.,  {Eisenstein} D.~J.,  {Hernquist} L.,  {Pillepich} A.,  {Nelson} D.,
   {Marinacci} F.,  {Springel} V.,   {Vogelsberger} M.,  2019, \mn@doi [\mnras]
  {10.1093/mnras/stz2546}, \href
  {https://ui.adsabs.harvard.edu/abs/2019MNRAS.tmp.2192B} {p.~2192}

\bibitem[\protect\citeauthoryear{{Campbell}, {van den Bosch}, {Padmanabhan},
  {Mao}, {Zentner}, {Lange}, {Jiang}  \& {Villarreal}}{{Campbell}
  et~al.}{2018}]{Campbell2018_sham}
{Campbell} D.,  {van den Bosch} F.~C.,  {Padmanabhan} N.,  {Mao} Y.-Y.,
  {Zentner} A.~R.,  {Lange} J.~U.,  {Jiang} F.,   {Villarreal} A.,  2018,
  \mn@doi [\mnras] {10.1093/mnras/sty495}, \href
  {https://ui.adsabs.harvard.edu/abs/2018MNRAS.477..359C} {477, 359}

\bibitem[\protect\citeauthoryear{{Chaves-Montero} \& {Hearin}}{{Chaves-Montero}
  \& {Hearin}}{2020}]{chaves_hearin2020_sbu1}
{Chaves-Montero} J.,  {Hearin} A.,  2020, \mn@doi [\mnras]
  {10.1093/mnras/staa1230}, \href
  {https://ui.adsabs.harvard.edu/abs/2020MNRAS.495.2088C} {495, 2088}

\bibitem[\protect\citeauthoryear{{Chaves-Montero}, {Angulo}, {Schaye},
  {Schaller}, {Crain}, {Furlong}  \& {Theuns}}{{Chaves-Montero}
  et~al.}{2016}]{ChavesMontero2016}
{Chaves-Montero} J.,  {Angulo} R.~E.,  {Schaye} J.,  {Schaller} M.,  {Crain}
  R.~A.,  {Furlong} M.,   {Theuns} T.,  2016, \mn@doi [\mnras]
  {10.1093/mnras/stw1225}, \href
  {https://ui.adsabs.harvard.edu/abs/2016MNRAS.460.3100C} {460, 3100}

\bibitem[\protect\citeauthoryear{{Chue}, {Dalal}  \& {White}}{{Chue}
  et~al.}{2018}]{Chue2018}
{Chue} C. Y.~R.,  {Dalal} N.,   {White} M.,  2018, \mn@doi [\jcap]
  {10.1088/1475-7516/2018/10/012}, \href
  {https://ui.adsabs.harvard.edu/abs/2018JCAP...10..012C} {2018, 012}

\bibitem[\protect\citeauthoryear{Conroy \& Wechsler}{Conroy \&
  Wechsler}{2009}]{conroy2009_CONNECTINGGALAXIESHALOS}
Conroy C.,  Wechsler R.~H.,  2009, \mn@doi [The Astrophysical Journal]
  {10.1088/0004-637X/696/1/620}, 696, 620

\bibitem[\protect\citeauthoryear{{Contreras}, {Angulo}  \&
  {Zennaro}}{{Contreras} et~al.}{2020}]{Contreras2020}
{Contreras} S.,  {Angulo} R.,   {Zennaro} M.,  2020, arXiv e-prints, \href
  {https://ui.adsabs.harvard.edu/abs/2020arXiv200503672C} {p. arXiv:2005.03672}

\bibitem[\protect\citeauthoryear{Cowie, Songaila, Hu  \& Cohen}{Cowie
  et~al.}{1996}]{cowie1996_NewInsightGalaxy}
Cowie L.~L.,  Songaila A.,  Hu E.~M.,   Cohen J.~G.,  1996, \mn@doi [\aj]
  {10.1086/118058}, 112, 839

\bibitem[\protect\citeauthoryear{{Dalal}, {White}, {Bond}  \&
  {Shirokov}}{{Dalal} et~al.}{2008}]{Dalal2008}
{Dalal} N.,  {White} M.,  {Bond} J.~R.,   {Shirokov} A.,  2008, \mn@doi [\apj]
  {10.1086/591512}, \href {http://adsabs.harvard.edu/abs/2008ApJ...687...12D}
  {687, 12}

\bibitem[\protect\citeauthoryear{Dav{\'e}, Thompson  \& Hopkins}{Dav{\'e}
  et~al.}{2016}]{dave2016_MUFASAGalaxyformation}
Dav{\'e} R.,  Thompson R.,   Hopkins P.~F.,  2016, \mn@doi [\mnras]
  {10.1093/mnras/stw1862}, 462, 3265

\bibitem[\protect\citeauthoryear{{Davis}, {Efstathiou}, {Frenk}  \&
  {White}}{{Davis} et~al.}{1985}]{Davis1985}
{Davis} M.,  {Efstathiou} G.,  {Frenk} C.~S.,   {White} S.~D.~M.,  1985,
  \mn@doi [\apj] {10.1086/163168}, \href
  {https://ui.adsabs.harvard.edu/abs/1985ApJ...292..371D} {292, 371}

\bibitem[\protect\citeauthoryear{{Dolag}, {Borgani}, {Murante}  \&
  {Springel}}{{Dolag} et~al.}{2009}]{Dolag2009}
{Dolag} K.,  {Borgani} S.,  {Murante} G.,   {Springel} V.,  2009, \mn@doi
  [\mnras] {10.1111/j.1365-2966.2009.15034.x}, \href
  {https://ui.adsabs.harvard.edu/abs/2009MNRAS.399..497D} {399, 497}

\bibitem[\protect\citeauthoryear{{Favole}, {Montero-Dorta}, {Artale},
  {Contreras}, {Zehavi}  \& {Xu}}{{Favole} et~al.}{2021}]{Favole2021}
{Favole} G.,  {Montero-Dorta} A.~D.,  {Artale} M.~C.,  {Contreras} S.,
  {Zehavi} I.,   {Xu} X.,  2021, arXiv e-prints, \href
  {https://ui.adsabs.harvard.edu/abs/2021arXiv210110733F} {p. arXiv:2101.10733}

\bibitem[\protect\citeauthoryear{Feldmann, Quataert, Hopkins,
  {Faucher-Gigu{\`e}re}  \& Kere{\v s}}{Feldmann
  et~al.}{2017}]{feldmann2017_ColoursStarformation}
Feldmann R.,  Quataert E.,  Hopkins P.~F.,  {Faucher-Gigu{\`e}re} C.-A.,
  Kere{\v s} D.,  2017, \mn@doi [Monthly Notices of the Royal Astronomical
  Society] {10.1093/mnras/stx1120}, 470, 1050

\bibitem[\protect\citeauthoryear{{Feldmann}, {Faucher-Gigu{\`e}re}  \&
  {Kere{\v{s}}}}{{Feldmann} et~al.}{2019}]{Feldmann2019}
{Feldmann} R.,  {Faucher-Gigu{\`e}re} C.-A.,   {Kere{\v{s}}} D.,  2019, \mn@doi
  [\apjl] {10.3847/2041-8213/aafe80}, \href
  {https://ui.adsabs.harvard.edu/abs/2019ApJ...871L..21F} {871, L21}

\bibitem[\protect\citeauthoryear{Fontanot, De~Lucia, Monaco, Somerville  \&
  Santini}{Fontanot et~al.}{2009}]{fontanot2009_ManyManifestationsdownsizing}
Fontanot F.,  De~Lucia G.,  Monaco P.,  Somerville R.~S.,   Santini P.,  2009,
  \mn@doi [\mnras] {10.1111/j.1365-2966.2009.15058.x}, 397, 1776

\bibitem[\protect\citeauthoryear{{Gao} \& {White}}{{Gao} \&
  {White}}{2007}]{Gao2007}
{Gao} L.,  {White} S.~D.~M.,  2007, \mn@doi [\mnras]
  {10.1111/j.1745-3933.2007.00292.x}, \href
  {http://adsabs.harvard.edu/abs/2007MNRAS.377L...5G} {377, L5}

\bibitem[\protect\citeauthoryear{{Gao}, {Springel}  \& {White}}{{Gao}
  et~al.}{2005}]{gao2005}
{Gao} L.,  {Springel} V.,   {White} S.~D.~M.,  2005, \mn@doi [\mnras]
  {10.1111/j.1745-3933.2005.00084.x}, \href
  {http://adsabs.harvard.edu/abs/2005MNRAS.363L..66G} {363, L66}

\bibitem[\protect\citeauthoryear{{Genel} et~al.,}{{Genel}
  et~al.}{2014}]{Genel2014}
{Genel} S.,  et~al., 2014, \mn@doi [\mnras] {10.1093/mnras/stu1654}, \href
  {https://ui.adsabs.harvard.edu/abs/2014MNRAS.445..175G} {445, 175}

\bibitem[\protect\citeauthoryear{{Gu}, {Conroy}  \& {Behroozi}}{{Gu}
  et~al.}{2016}]{Gu2016}
{Gu} M.,  {Conroy} C.,   {Behroozi} P.,  2016, \mn@doi [\apj]
  {10.3847/0004-637X/833/1/2}, \href
  {https://ui.adsabs.harvard.edu/abs/2016ApJ...833....2G} {833, 2}

\bibitem[\protect\citeauthoryear{{Gu} et~al.,}{{Gu} et~al.}{2020}]{Gu2020}
{Gu} M.,  et~al., 2020, arXiv e-prints, \href
  {https://ui.adsabs.harvard.edu/abs/2020arXiv201004166G} {p. arXiv:2010.04166}

\bibitem[\protect\citeauthoryear{{Hadzhiyska}, {Bose}, {Eisenstein},
  {Hernquist}  \& {Spergel}}{{Hadzhiyska} et~al.}{2020}]{Hadzhiyska2020}
{Hadzhiyska} B.,  {Bose} S.,  {Eisenstein} D.,  {Hernquist} L.,   {Spergel}
  D.~N.,  2020, \mn@doi [\mnras] {10.1093/mnras/staa623}, \href
  {https://ui.adsabs.harvard.edu/abs/2020MNRAS.493.5506H} {493, 5506}

\bibitem[\protect\citeauthoryear{{Hadzhiyska}, {Bose}, {Eisenstein}  \&
  {Hernquist}}{{Hadzhiyska} et~al.}{2021a}]{Hadzhiyska2021_sham}
{Hadzhiyska} B.,  {Bose} S.,  {Eisenstein} D.,   {Hernquist} L.,  2021a,
  \mn@doi [\mnras] {10.1093/mnras/staa3776}, \href
  {https://ui.adsabs.harvard.edu/abs/2021MNRAS.501.1603H} {501, 1603}

\bibitem[\protect\citeauthoryear{{Hadzhiyska}, {Bose}, {Eisenstein}  \&
  {Hernquist}}{{Hadzhiyska} et~al.}{2021b}]{Hadzhiyska2021}
{Hadzhiyska} B.,  {Bose} S.,  {Eisenstein} D.,   {Hernquist} L.,  2021b,
  \mn@doi [\mnras] {10.1093/mnras/staa3776}, \href
  {https://ui.adsabs.harvard.edu/abs/2021MNRAS.501.1603H} {501, 1603}

\bibitem[\protect\citeauthoryear{{Han}, {Li}, {Jing}, {Nishimichi}, {Wang}  \&
  {Jiang}}{{Han} et~al.}{2019}]{han2018}
{Han} J.,  {Li} Y.,  {Jing} Y.,  {Nishimichi} T.,  {Wang} W.,   {Jiang} C.,
  2019, \mn@doi [\mnras] {10.1093/mnras/sty2822}, \href
  {https://ui.adsabs.harvard.edu/abs/2019MNRAS.482.1900H} {482, 1900}

\bibitem[\protect\citeauthoryear{{Hearin} \& {Watson}}{{Hearin} \&
  {Watson}}{2013}]{Hearin2013}
{Hearin} A.~P.,  {Watson} D.~F.,  2013, \mn@doi [\mnras]
  {10.1093/mnras/stt1374}, \href
  {http://adsabs.harvard.edu/abs/2013MNRAS.435.1313H} {435, 1313}

\bibitem[\protect\citeauthoryear{{Hearin}, {Zentner}, {van den Bosch},
  {Campbell}  \& {Tollerud}}{{Hearin} et~al.}{2016}]{Hearin2016}
{Hearin} A.~P.,  {Zentner} A.~R.,  {van den Bosch} F.~C.,  {Campbell} D.,
  {Tollerud} E.,  2016, \mn@doi [\mnras] {10.1093/mnras/stw840}, \href
  {http://adsabs.harvard.edu/abs/2016MNRAS.460.2552H} {460, 2552}

\bibitem[\protect\citeauthoryear{{Johnson}, {Maller}, {Berlind}, {Sinha}  \&
  {Holley-Bockelmann}}{{Johnson} et~al.}{2019}]{Johnson2019}
{Johnson} J.~W.,  {Maller} A.~H.,  {Berlind} A.~A.,  {Sinha} M.,
  {Holley-Bockelmann} J.~K.,  2019, \mn@doi [\mnras] {10.1093/mnras/stz942},
  \href {https://ui.adsabs.harvard.edu/abs/2019MNRAS.486.1156J} {486, 1156}

\bibitem[\protect\citeauthoryear{{Kauffmann}, {White}  \&
  {Guiderdoni}}{{Kauffmann} et~al.}{1993}]{Kauffmann1993}
{Kauffmann} G.,  {White} S.~D.~M.,   {Guiderdoni} B.,  1993, \mnras, \href
  {http://adsabs.harvard.edu/abs/1993MNRAS.264..201K} {264, 201}

\bibitem[\protect\citeauthoryear{{Knebe} et~al.,}{{Knebe}
  et~al.}{2011}]{knebe2011_comparison}
{Knebe} A.,  et~al., 2011, \mn@doi [\mnras] {10.1111/j.1365-2966.2011.18858.x},
  \href {https://ui.adsabs.harvard.edu/abs/2011MNRAS.415.2293K} {415, 2293}

\bibitem[\protect\citeauthoryear{{Knebe} et~al.,}{{Knebe}
  et~al.}{2013}]{Knebe2013_comparison}
{Knebe} A.,  et~al., 2013, \mn@doi [\mnras] {10.1093/mnras/stt1403}, \href
  {https://ui.adsabs.harvard.edu/abs/2013MNRAS.435.1618K} {435, 1618}

\bibitem[\protect\citeauthoryear{{Landy} \& {Szalay}}{{Landy} \&
  {Szalay}}{1993}]{Landy1993}
{Landy} S.~D.,  {Szalay} A.~S.,  1993, \mn@doi [\apj] {10.1086/172900}, \href
  {http://adsabs.harvard.edu/abs/1993ApJ...412...64L} {412, 64}

\bibitem[\protect\citeauthoryear{{Lazeyras}, {Musso}  \& {Schmidt}}{{Lazeyras}
  et~al.}{2017}]{Lazeyras2017}
{Lazeyras} T.,  {Musso} M.,   {Schmidt} F.,  2017, \mn@doi [\jcap]
  {10.1088/1475-7516/2017/03/059}, \href
  {https://ui.adsabs.harvard.edu/abs/2017JCAP...03..059L} {2017, 059}

\bibitem[\protect\citeauthoryear{{Li}, {Mo}  \& {Gao}}{{Li}
  et~al.}{2008}]{2008Li}
{Li} Y.,  {Mo} H.~J.,   {Gao} L.,  2008, \mn@doi [\mnras]
  {10.1111/j.1365-2966.2008.13667.x}, \href
  {http://adsabs.harvard.edu/abs/2008MNRAS.389.1419L} {389, 1419}

\bibitem[\protect\citeauthoryear{{Lin}, {Mandelbaum}, {Huang}, {Huang},
  {Dalal}, {Diemer}, {Jian}  \& {Kravtsov}}{{Lin} et~al.}{2016}]{Lin2016}
{Lin} Y.-T.,  {Mandelbaum} R.,  {Huang} Y.-H.,  {Huang} H.-J.,  {Dalal} N.,
  {Diemer} B.,  {Jian} H.-Y.,   {Kravtsov} A.,  2016, \mn@doi [\apj]
  {10.3847/0004-637X/819/2/119}, \href
  {http://adsabs.harvard.edu/abs/2016ApJ...819..119L} {819, 119}

\bibitem[\protect\citeauthoryear{{Mao}, {Zentner}  \& {Wechsler}}{{Mao}
  et~al.}{2018}]{Mao2018}
{Mao} Y.-Y.,  {Zentner} A.~R.,   {Wechsler} R.~H.,  2018, \mn@doi [\mnras]
  {10.1093/mnras/stx3111}, \href
  {https://ui.adsabs.harvard.edu/abs/2018MNRAS.474.5143M} {474, 5143}

\bibitem[\protect\citeauthoryear{{Marinacci} et~al.,}{{Marinacci}
  et~al.}{2018}]{Marinacci2018}
{Marinacci} F.,  et~al., 2018, \mn@doi [\mnras] {10.1093/mnras/sty2206}, \href
  {https://ui.adsabs.harvard.edu/abs/2018MNRAS.480.5113M} {480, 5113}

\bibitem[\protect\citeauthoryear{{Miyatake}, {More}, {Takada}, {Spergel},
  {Mandelbaum}, {Rykoff}  \& {Rozo}}{{Miyatake} et~al.}{2016}]{Miyatake2016}
{Miyatake} H.,  {More} S.,  {Takada} M.,  {Spergel} D.~N.,  {Mandelbaum} R.,
  {Rykoff} E.~S.,   {Rozo} E.,  2016, \mn@doi [Physical Review Letters]
  {10.1103/PhysRevLett.116.041301}, \href
  {http://adsabs.harvard.edu/abs/2016PhRvL.116d1301M} {116, 041301}

\bibitem[\protect\citeauthoryear{{Montero-Dorta} et~al.,}{{Montero-Dorta}
  et~al.}{2017}]{MonteroDorta2017B}
{Montero-Dorta} A.~D.,  et~al., 2017, \mn@doi [\apjl]
  {10.3847/2041-8213/aa8cc5}, \href
  {http://adsabs.harvard.edu/abs/2017ApJ...848L...2M} {848, L2}

\bibitem[\protect\citeauthoryear{{Montero-Dorta}, {Artale}, {Abramo}  \&
  {Tucci}}{{Montero-Dorta} et~al.}{2020a}]{MonteroDorta2020C}
{Montero-Dorta} A.~D.,  {Artale} M.~C.,  {Abramo} L.~R.,   {Tucci} B.,  2020a,
  arXiv e-prints, \href {https://ui.adsabs.harvard.edu/abs/2020arXiv200808607M}
  {p. arXiv:2008.08607}

\bibitem[\protect\citeauthoryear{{Montero-Dorta} et~al.,}{{Montero-Dorta}
  et~al.}{2020b}]{MonteroDorta2020B}
{Montero-Dorta} A.~D.,  et~al., 2020b, \mn@doi [\mnras]
  {10.1093/mnras/staa1624}, \href
  {https://ui.adsabs.harvard.edu/abs/2020MNRAS.496.1182M} {496, 1182}

\bibitem[\protect\citeauthoryear{{More}, {van den Bosch}, {Cacciato}, {Mo},
  {Yang}  \& {Li}}{{More} et~al.}{2009}]{More2009}
{More} S.,  {van den Bosch} F.~C.,  {Cacciato} M.,  {Mo} H.~J.,  {Yang} X.,
  {Li} R.,  2009, \mn@doi [\mnras] {10.1111/j.1365-2966.2008.14095.x}, \href
  {https://ui.adsabs.harvard.edu/abs/2009MNRAS.392..801M} {392, 801}

\bibitem[\protect\citeauthoryear{{Moster}, {Naab}  \& {White}}{{Moster}
  et~al.}{2018}]{moster2018_emerge}
{Moster} B.~P.,  {Naab} T.,   {White} S. D.~M.,  2018, \mn@doi [\mnras]
  {10.1093/mnras/sty655}, \href
  {https://ui.adsabs.harvard.edu/abs/2018MNRAS.477.1822M} {477, 1822}

\bibitem[\protect\citeauthoryear{{Musso}, {Cadiou}, {Pichon}, {Codis},
  {Kraljic}  \& {Dubois}}{{Musso} et~al.}{2018}]{Musso2018}
{Musso} M.,  {Cadiou} C.,  {Pichon} C.,  {Codis} S.,  {Kraljic} K.,   {Dubois}
  Y.,  2018, \mn@doi [\mnras] {10.1093/mnras/sty191}, \href
  {https://ui.adsabs.harvard.edu/abs/2018MNRAS.476.4877M} {476, 4877}

\bibitem[\protect\citeauthoryear{{Naab} \& {Ostriker}}{{Naab} \&
  {Ostriker}}{2017}]{Naab2017}
{Naab} T.,  {Ostriker} J.~P.,  2017, \mn@doi [\araa]
  {10.1146/annurev-astro-081913-040019}, \href
  {https://ui.adsabs.harvard.edu/abs/2017ARA&A..55...59N} {55, 59}

\bibitem[\protect\citeauthoryear{{Naiman} et~al.,}{{Naiman}
  et~al.}{2018}]{Naiman2018}
{Naiman} J.~P.,  et~al., 2018, \mn@doi [\mnras] {10.1093/mnras/sty618}, \href
  {https://ui.adsabs.harvard.edu/abs/2018MNRAS.477.1206N} {477, 1206}

\bibitem[\protect\citeauthoryear{{Nelson} et~al.,}{{Nelson}
  et~al.}{2018a}]{Nelson2018_ColorBim}
{Nelson} D.,  et~al., 2018a, \mn@doi [\mnras] {10.1093/mnras/stx3040}, \href
  {https://ui.adsabs.harvard.edu/abs/2018MNRAS.475..624N} {475, 624}

\bibitem[\protect\citeauthoryear{{Nelson} et~al.,}{{Nelson}
  et~al.}{2018b}]{Nelson2018}
{Nelson} D.,  et~al., 2018b, \mn@doi [\mnras] {10.1093/mnras/stx3040}, \href
  {https://ui.adsabs.harvard.edu/abs/2018MNRAS.475..624N} {475, 624}

\bibitem[\protect\citeauthoryear{{Nelson} et~al.,}{{Nelson}
  et~al.}{2019}]{Nelson2019}
{Nelson} D.,  et~al., 2019, \mn@doi [Computational Astrophysics and Cosmology]
  {10.1186/s40668-019-0028-x}, \href
  {https://ui.adsabs.harvard.edu/abs/2019ComAC...6....2N} {6, 2}

\bibitem[\protect\citeauthoryear{{Niemiec} et~al.,}{{Niemiec}
  et~al.}{2018}]{Niemiec2018}
{Niemiec} A.,  et~al., 2018, \mn@doi [\mnras] {10.1093/mnrasl/sly041}, \href
  {http://adsabs.harvard.edu/abs/2018MNRAS.tmpL..42N} {}

\bibitem[\protect\citeauthoryear{{Obuljen}, {Dalal}  \& {Percival}}{{Obuljen}
  et~al.}{2019}]{Obuljen2019}
{Obuljen} A.,  {Dalal} N.,   {Percival} W.~J.,  2019, \mn@doi [\jcap]
  {10.1088/1475-7516/2019/10/020}, \href
  {https://ui.adsabs.harvard.edu/abs/2019JCAP...10..020O} {2019, 020}

\bibitem[\protect\citeauthoryear{{Obuljen}, {Percival}  \& {Dalal}}{{Obuljen}
  et~al.}{2020}]{Obuljen2020}
{Obuljen} A.,  {Percival} W.~J.,   {Dalal} N.,  2020, \mn@doi [\jcap]
  {10.1088/1475-7516/2020/10/058}, \href
  {https://ui.adsabs.harvard.edu/abs/2020JCAP...10..058O} {2020, 058}

\bibitem[\protect\citeauthoryear{{Pillepich} et~al.,}{{Pillepich}
  et~al.}{2018a}]{Pillepich2018}
{Pillepich} A.,  et~al., 2018a, \mn@doi [\mnras] {10.1093/mnras/stx2656}, \href
  {https://ui.adsabs.harvard.edu/abs/2018MNRAS.473.4077P} {473, 4077}

\bibitem[\protect\citeauthoryear{{Pillepich} et~al.,}{{Pillepich}
  et~al.}{2018b}]{Pillepich2018b}
{Pillepich} A.,  et~al., 2018b, \mn@doi [\mnras] {10.1093/mnras/stx3112}, \href
  {https://ui.adsabs.harvard.edu/abs/2018MNRAS.475..648P} {475, 648}

\bibitem[\protect\citeauthoryear{{Planck Collaboration} et~al.,}{{Planck
  Collaboration} et~al.}{2016}]{Planck2016}
{Planck Collaboration} et~al., 2016, \mn@doi [\aap]
  {10.1051/0004-6361/201525830}, \href
  {https://ui.adsabs.harvard.edu/abs/2016A&A...594A..13P} {594, A13}

\bibitem[\protect\citeauthoryear{{Press} \& {Schechter}}{{Press} \&
  {Schechter}}{1974}]{Press1974}
{Press} W.~H.,  {Schechter} P.,  1974, \mn@doi [\apj] {10.1086/152650}, \href
  {http://adsabs.harvard.edu/abs/1974ApJ...187..425P} {187, 425}

\bibitem[\protect\citeauthoryear{{Ramakrishnan}, {Paranjape}, {Hahn}  \&
  {Sheth}}{{Ramakrishnan} et~al.}{2019}]{Ramakrishnan2019}
{Ramakrishnan} S.,  {Paranjape} A.,  {Hahn} O.,   {Sheth} R.~K.,  2019, \mn@doi
  [\mnras] {10.1093/mnras/stz2344}, \href
  {https://ui.adsabs.harvard.edu/abs/2019MNRAS.489.2977R} {489, 2977}

\bibitem[\protect\citeauthoryear{{Reddick}, {Wechsler}, {Tinker}  \&
  {Behroozi}}{{Reddick} et~al.}{2013}]{Reddick2013}
{Reddick} R.~M.,  {Wechsler} R.~H.,  {Tinker} J.~L.,   {Behroozi} P.~S.,  2013,
  \mn@doi [\apj] {10.1088/0004-637X/771/1/30}, \href
  {https://ui.adsabs.harvard.edu/abs/2013ApJ...771...30R} {771, 30}

\bibitem[\protect\citeauthoryear{{Rodriguez-Gomez} et~al.,}{{Rodriguez-Gomez}
  et~al.}{2015}]{Rodriguez-Gomez2015}
{Rodriguez-Gomez} V.,  et~al., 2015, \mn@doi [\mnras] {10.1093/mnras/stv264},
  \href {https://ui.adsabs.harvard.edu/abs/2015MNRAS.449...49R} {449, 49}

\bibitem[\protect\citeauthoryear{{Salcedo}, {Maller}, {Berlind}, {Sinha},
  {McBride}, {Behroozi}, {Wechsler}  \& {Weinberg}}{{Salcedo}
  et~al.}{2018}]{2018Salcedo}
{Salcedo} A.~N.,  {Maller} A.~H.,  {Berlind} A.~A.,  {Sinha} M.,  {McBride}
  C.~K.,  {Behroozi} P.~S.,  {Wechsler} R.~H.,   {Weinberg} D.~H.,  2018,
  \mn@doi [\mnras] {10.1093/mnras/sty109}, \href
  {https://ui.adsabs.harvard.edu/abs/2018MNRAS.475.4411S} {475, 4411}

\bibitem[\protect\citeauthoryear{{Salcedo} et~al.,}{{Salcedo}
  et~al.}{2020}]{Salcedo2020}
{Salcedo} A.~N.,  et~al., 2020, arXiv e-prints, \href
  {https://ui.adsabs.harvard.edu/abs/2020arXiv201004176S} {p. arXiv:2010.04176}

\bibitem[\protect\citeauthoryear{{Sato-Polito}, {Montero-Dorta}, {Abramo},
  {Prada}  \& {Klypin}}{{Sato-Polito} et~al.}{2019}]{SatoPolito2019}
{Sato-Polito} G.,  {Montero-Dorta} A.~D.,  {Abramo} L.~R.,  {Prada} F.,
  {Klypin} A.,  2019, \mn@doi [\mnras] {10.1093/mnras/stz1338}, \href
  {https://ui.adsabs.harvard.edu/abs/2019MNRAS.487.1570S} {487, 1570}

\bibitem[\protect\citeauthoryear{{Sheth} \& {Tormen}}{{Sheth} \&
  {Tormen}}{2002}]{ShethTormen2002}
{Sheth} R.~K.,  {Tormen} G.,  2002, \mn@doi [\mnras]
  {10.1046/j.1365-8711.2002.04950.x}, \href
  {http://adsabs.harvard.edu/abs/2002MNRAS.329...61S} {329, 61}

\bibitem[\protect\citeauthoryear{{Sheth} \& {Tormen}}{{Sheth} \&
  {Tormen}}{2004}]{Sheth2004}
{Sheth} R.~K.,  {Tormen} G.,  2004, \mn@doi [\mnras]
  {10.1111/j.1365-2966.2004.07733.x}, \href
  {https://ui.adsabs.harvard.edu/abs/2004MNRAS.350.1385S} {350, 1385}

\bibitem[\protect\citeauthoryear{{Shi} et~al.,}{{Shi} et~al.}{2020}]{Shi2020}
{Shi} J.,  et~al., 2020, \mn@doi [\apj] {10.3847/1538-4357/ab8464}, \href
  {https://ui.adsabs.harvard.edu/abs/2020ApJ...893..139S} {893, 139}

\bibitem[\protect\citeauthoryear{{Somerville} \& {Dav{\'e}}}{{Somerville} \&
  {Dav{\'e}}}{2015}]{Somerville2015}
{Somerville} R.~S.,  {Dav{\'e}} R.,  2015, \mn@doi [\araa]
  {10.1146/annurev-astro-082812-140951}, \href
  {https://ui.adsabs.harvard.edu/abs/2015ARA&A..53...51S} {53, 51}

\bibitem[\protect\citeauthoryear{{Springel}}{{Springel}}{2010}]{Springel2010}
{Springel} V.,  2010, \mn@doi [\mnras] {10.1111/j.1365-2966.2009.15715.x},
  \href {https://ui.adsabs.harvard.edu/abs/2010MNRAS.401..791S} {401, 791}

\bibitem[\protect\citeauthoryear{{Springel}, {White}, {Tormen}  \&
  {Kauffmann}}{{Springel} et~al.}{2001}]{Springel2001}
{Springel} V.,  {White} S. D.~M.,  {Tormen} G.,   {Kauffmann} G.,  2001,
  \mn@doi [\mnras] {10.1046/j.1365-8711.2001.04912.x}, \href
  {https://ui.adsabs.harvard.edu/abs/2001MNRAS.328..726S} {328, 726}

\bibitem[\protect\citeauthoryear{{Springel} et~al.,}{{Springel}
  et~al.}{2005}]{Springel2005}
{Springel} V.,  et~al., 2005, \mn@doi [\nat] {10.1038/nature03597}, \href
  {http://adsabs.harvard.edu/abs/2005Natur.435..629S} {435, 629}

\bibitem[\protect\citeauthoryear{{Springel} et~al.,}{{Springel}
  et~al.}{2018}]{Springel2018}
{Springel} V.,  et~al., 2018, \mn@doi [\mnras] {10.1093/mnras/stx3304}, \href
  {https://ui.adsabs.harvard.edu/abs/2018MNRAS.475..676S} {475, 676}

\bibitem[\protect\citeauthoryear{{Sunayama} \& {More}}{{Sunayama} \&
  {More}}{2019}]{Sunayama2019}
{Sunayama} T.,  {More} S.,  2019, \mn@doi [\mnras] {10.1093/mnras/stz2832},
  \href {https://ui.adsabs.harvard.edu/abs/2019MNRAS.490.4945S} {490, 4945}

\bibitem[\protect\citeauthoryear{{Sunayama}, {Hearin}, {Padmanabhan}  \&
  {Leauthaud}}{{Sunayama} et~al.}{2016}]{Sunayama2016}
{Sunayama} T.,  {Hearin} A.~P.,  {Padmanabhan} N.,   {Leauthaud} A.,  2016,
  \mn@doi [\mnras] {10.1093/mnras/stw332}, \href
  {http://adsabs.harvard.edu/abs/2016MNRAS.458.1510S} {458, 1510}

\bibitem[\protect\citeauthoryear{{Tinker}, {Leauthaud}, {Bundy}, {George},
  {Behroozi}, {Massey}, {Rhodes}  \& {Wechsler}}{{Tinker}
  et~al.}{2013}]{Tinker2013}
{Tinker} J.~L.,  {Leauthaud} A.,  {Bundy} K.,  {George} M.~R.,  {Behroozi} P.,
  {Massey} R.,  {Rhodes} J.,   {Wechsler} R.~H.,  2013, \mn@doi [\apj]
  {10.1088/0004-637X/778/2/93}, \href
  {https://ui.adsabs.harvard.edu/abs/2013ApJ...778...93T} {778, 93}

\bibitem[\protect\citeauthoryear{{Tucci}, {Montero-Dorta}, {Abramo},
  {Sato-Polito}  \& {Artale}}{{Tucci} et~al.}{2021}]{Tucci2020}
{Tucci} B.,  {Montero-Dorta} A.~D.,  {Abramo} L.~R.,  {Sato-Polito} G.,
  {Artale} M.~C.,  2021, \mn@doi [\mnras] {10.1093/mnras/staa3319}, \href
  {https://ui.adsabs.harvard.edu/abs/2021MNRAS.500.2777T} {500, 2777}

\bibitem[\protect\citeauthoryear{{Vogelsberger} et~al.,}{{Vogelsberger}
  et~al.}{2014a}]{Vogelsberger2014a}
{Vogelsberger} M.,  et~al., 2014a, \mn@doi [\mnras] {10.1093/mnras/stu1536},
  \href {https://ui.adsabs.harvard.edu/abs/2014MNRAS.444.1518V} {444, 1518}

\bibitem[\protect\citeauthoryear{{Vogelsberger} et~al.,}{{Vogelsberger}
  et~al.}{2014b}]{Vogelsberger2014b}
{Vogelsberger} M.,  et~al., 2014b, \mn@doi [\nat] {10.1038/nature13316}, \href
  {https://ui.adsabs.harvard.edu/abs/2014Natur.509..177V} {509, 177}

\bibitem[\protect\citeauthoryear{{Walsh} \& {Tinker}}{{Walsh} \&
  {Tinker}}{2019}]{Walsh2019}
{Walsh} K.,  {Tinker} J.,  2019, \mn@doi [\mnras] {10.1093/mnras/stz1351},
  \href {https://ui.adsabs.harvard.edu/abs/2019MNRAS.488..470W} {488, 470}

\bibitem[\protect\citeauthoryear{{Wang}, {Mao}, {Zentner}, {Lange}, {van den
  Bosch}  \& {Wechsler}}{{Wang} et~al.}{2020}]{Wang2020_concentration}
{Wang} K.,  {Mao} Y.-Y.,  {Zentner} A.~R.,  {Lange} J.~U.,  {van den Bosch}
  F.~C.,   {Wechsler} R.~H.,  2020, \mn@doi [\mnras] {10.1093/mnras/staa2733},
  \href {https://ui.adsabs.harvard.edu/abs/2020MNRAS.498.4450W} {498, 4450}

\bibitem[\protect\citeauthoryear{{Watson} \& {Conroy}}{{Watson} \&
  {Conroy}}{2013}]{Watson2013}
{Watson} D.~F.,  {Conroy} C.,  2013, \mn@doi [\apj]
  {10.1088/0004-637X/772/2/139}, \href
  {https://ui.adsabs.harvard.edu/abs/2013ApJ...772..139W} {772, 139}

\bibitem[\protect\citeauthoryear{{Wechsler} \& {Tinker}}{{Wechsler} \&
  {Tinker}}{2018}]{Wechsler2018}
{Wechsler} R.~H.,  {Tinker} J.~L.,  2018, \mn@doi [\araa]
  {10.1146/annurev-astro-081817-051756}, \href
  {https://ui.adsabs.harvard.edu/abs/2018ARA&A..56..435W} {56, 435}

\bibitem[\protect\citeauthoryear{{Wechsler}, {Zentner}, {Bullock}, {Kravtsov}
  \& {Allgood}}{{Wechsler} et~al.}{2006}]{wechsler2006}
{Wechsler} R.~H.,  {Zentner} A.~R.,  {Bullock} J.~S.,  {Kravtsov} A.~V.,
  {Allgood} B.,  2006, \mn@doi [\apj] {10.1086/507120}, \href
  {http://adsabs.harvard.edu/abs/2006ApJ...652...71W} {652, 71}

\bibitem[\protect\citeauthoryear{{White} \& {Rees}}{{White} \&
  {Rees}}{1978}]{White1978}
{White} S.~D.~M.,  {Rees} M.~J.,  1978, \mnras, \href
  {http://adsabs.harvard.edu/abs/1978MNRAS.183..341W} {183, 341}

\bibitem[\protect\citeauthoryear{{Xu}, {Zehavi}  \& {Contreras}}{{Xu}
  et~al.}{2021}]{Xu2021}
{Xu} X.,  {Zehavi} I.,   {Contreras} S.,  2021, \mn@doi [\mnras]
  {10.1093/mnras/stab100}, \href
  {https://ui.adsabs.harvard.edu/abs/2021MNRAS.502.3242X} {502, 3242}

\bibitem[\protect\citeauthoryear{{Yang}, {Mo}  \& {van den Bosch}}{{Yang}
  et~al.}{2009}]{Yang2009}
{Yang} X.,  {Mo} H.~J.,   {van den Bosch} F.~C.,  2009, \mn@doi [\apj]
  {10.1088/0004-637X/693/1/830}, \href
  {https://ui.adsabs.harvard.edu/abs/2009ApJ...693..830Y} {693, 830}

\bibitem[\protect\citeauthoryear{{Zehavi}, {Contreras}, {Padilla}, {Smith},
  {Baugh}  \& {Norberg}}{{Zehavi} et~al.}{2018}]{Zehavi2018}
{Zehavi} I.,  {Contreras} S.,  {Padilla} N.,  {Smith} N.~J.,  {Baugh} C.~M.,
  {Norberg} P.,  2018, \mn@doi [\apj] {10.3847/1538-4357/aaa54a}, \href
  {https://ui.adsabs.harvard.edu/abs/2018ApJ...853...84Z} {853, 84}

\bibitem[\protect\citeauthoryear{{Zentner}, {Hearin}, {van den Bosch}, {Lange}
  \& {Villarreal}}{{Zentner} et~al.}{2019}]{Zentner2019}
{Zentner} A.~R.,  {Hearin} A.,  {van den Bosch} F.~C.,  {Lange} J.~U.,
  {Villarreal} A.,  2019, \mn@doi [\mnras] {10.1093/mnras/stz470}, \href
  {https://ui.adsabs.harvard.edu/abs/2019MNRAS.485.1196Z} {485, 1196}

\bibitem[\protect\citeauthoryear{{Zu}, {Mandelbaum}, {Simet}, {Rozo}  \&
  {Rykoff}}{{Zu} et~al.}{2017}]{Zu2016}
{Zu} Y.,  {Mandelbaum} R.,  {Simet} M.,  {Rozo} E.,   {Rykoff} E.~S.,  2017,
  \mn@doi [\mnras] {10.1093/mnras/stx1264}, \href
  {https://ui.adsabs.harvard.edu/abs/2017MNRAS.470..551Z} {470, 551}

\makeatother
\end{thebibliography}

\label{lastpage}

\end{document}